\documentclass[11pt]{article}
\pdfoutput=1

\usepackage{amsmath,amssymb,exscale}
\usepackage{array,multicol}
\usepackage{afterpage,float,flafter}
\usepackage{booktabs,multirow}
\usepackage{epsfig,rotating,pifont}
\usepackage[sort]{cite}
\usepackage{color}
\usepackage{marvosym}
\usepackage{textcomp}
\usepackage{wrapfig}
\usepackage{hyperref}
\usepackage[all]{hypcap}

\definecolor{nicered}{rgb}{0.5,0.1,0.1}
\definecolor{nicegreen}{rgb}{0.1,0.5,0.1}
\definecolor{niceblue}{rgb}{0.1,0.1,0.8}
\hypersetup{colorlinks,citecolor=nicegreen,linkcolor=nicered}

\@addtoreset{equation}{section} \makeatother

\def\lsim{\mathrel{\rlap{\lower3pt\hbox{\hskip0pt$\sim$}}\raise1pt\hbox{$<$}}} 
\def\gsim{\mathrel{\rlap{\lower4pt\hbox{\hskip1pt$\sim$}}\raise1pt\hbox{$>$}}} 

\newcommand{\dslash}[1]{#1 \! \! \! {\bf /}}
\newcommand{\ddslash}[1]{#1 \! \! \! \! {\bf /}}

\title{
	\vspace{-3cm}
	\begin{flushright}
		\small{DESY 13-080}
	\end{flushright}
	\vspace{2.7cm}
	\begin{center}
		\medskip
		{ \Huge \bf Strong Signatures \\ \vspace{1mm} of Right-Handed Compositeness }
	\end{center}
	\vspace{0.6cm}
}
\author{
	\large{
		\text{\bf Michele Redi$^{1}$} \footnote{michele.redi@fi.infn.it} ~~
		\text{\bf Veronica Sanz$^{2,3}$} \footnote{veronica.sanz@cern.ch} ~~
		\text{\bf Maikel de Vries$^{4}$} \footnote{maikel.devries@desy.de} ~~
		\text{\bf Andreas Weiler$^{4}$} \footnote{andreas.weiler@desy.de}
	} \\ \\
	$^1$ \emph{INFN, Sezione di Firenze, Via G. Sansone 1, 50019 Sesto Fiorentino, Italy} \\
	$^2$ \emph{Deptartment of Physics and Astronomy, York University,} \\ \emph{4700 Keele Street, Toronto, Ontario, M3J 1P3, Canada} \\
	$^3$ \emph{Deptartment of Physics and Astronomy, University of Sussex,} \\ \emph{Sussex House, Brighton BN1 9QH, United Kingdom} \\
	$^4$ \emph{DESY, Notkestrasse 85, D-22607 Hamburg, Germany}
}
\date{}

\begin{document}

\maketitle \thispagestyle{empty} \vspace*{-.2cm}

\begin{abstract} \noindent
Right-handed light quarks could be significantly composite, yet compatible with experimental searches at the LHC and precision tests on Standard Model couplings. In these scenarios, that are motivated by flavor physics, one expects large cross sections for the production of new resonances coupled to light quarks. We study experimental strong signatures of right-handed compositeness at the LHC, and constrain the parameter space of these models with recent results by ATLAS and CMS. We show that the LHC sensitivity could be significantly improved if dedicated searches were performed, in particular in multi-jet signals.
\end{abstract}

\newpage
\renewcommand{\thepage}{\arabic{page}}
\setcounter{page}{1}

\section{Introduction}
\label{sec:introduction}

Modern realizations of composite Higgs models rely on the hypothesis of partial compositeness, each SM state has a heavy partner with equal quantum numbers under the SM symmetries, see \cite{bib:minimalcomposite,bib:reviewcomp} and references therein. Until recently most studies focused on the so called ``anarchic scenario'' where the SM light quarks are mostly elementary and the top largely composite \cite{bib:anarchic}. This hypothesis hides strong coupling effects from flavor and electroweak observables but also eliminates the typical collider signatures of compositeness.

In references \cite{Cacciapaglia:2007fw,bib:compositeMFV,bib:perez,bib:duccio} it was shown that a different philosophy is possible within the partial compositeness paradigm, where one chirality of SM light quarks has large compositeness. These scenarios are in fact strongly motivated by flavor physics. Assuming universal couplings for either left-handed or right-handed fermions allows to realize the hypothesis of Minimal Flavor Violation (MFV) \cite{bib:MFV} in strongly coupled theories, solving the flavor problem of composite Higgs models \cite{bib:flavorconstraints}. Here the compositeness of the up quark cannot be small, it being determined by the one of the top. Generalisations allowing to split the third generation can also be considered \cite{bib:su2}.

In this note we will focus on the phenomenologically attractive scenario of composite right-handed quarks that is weakly constrained by precision electroweak tests allowing a large degree of compositeness, see \cite{bib:Barbieri2012tu} for a recent discussion. We will study in detail the collider phenomenology extending and updating the results in \cite{bib:compositeMFV}. The experimental signatures are dramatically different from the ones of the widely studied anarchic models \cite{bib:lhcanarchic}. There the fact that the proton constituents are elementary makes it difficult to produce the new states at the LHC. If right-handed up and down quarks are composite instead, the couplings to the strong sector will be large. This implies larger production cross sections for the heavy states that can be tested with present LHC data.

The typical collider signatures of our scenario are jet final states. In particular we derive a strong bound on gluon resonances from the latest dijet searches at LHC. The phenomenology of heavy fermions depends on the chirality of the associated SM particles. Partners of left-handed quarks can be singly produced through electroweak interactions with large cross sections already at the 8 TeV LHC. This places a stringent and rather model independent bound  that can be extracted from an ATLAS search \cite{bib:ATLAS:2012apa}. Partners of right-handed quarks are instead more difficult to produce and lead to final states with up to six jets and no missing energy. We find that present multi-jet LHC searches, tailored for supersymmetric scenarios, are mostly insensitive to this signature even in the R-parity violation case. Bounds could be here significantly improved with dedicated searches and we suggest some possibilities that could be explored by the experimental collaborations.

The paper is organized as follows: In section \ref{sec:compositelightquarks} we review the model and discuss the relevant features of right-handed compositeness. We emphasize in particular the importance of chromomagnetic interactions. In section \ref{sec:coloroctet} we discuss the phenomenology of the color octet. The relevant experimental searches will be discussed and limits on the octet mass extracted. In sections \ref{sec:lefthandedpartners} and \ref{sec:righthandedpartners} the collider signatures of heavy quark partners will be discussed. Available searches will be analyzed and dedicated search strategies will be proposed in section \ref{sec:dedicatedsearches}. We conclude in section \ref{sec:conclusions}. In appendix \ref{sec:compositemodel} the model used in our simulations is presented and in appendix \ref{sec:ptordering} the $p_T$ distribution in single production of heavy quark partners is discussed.

\section{Composite Light Quarks}
\label{sec:compositelightquarks}
Within the framework of partial compositeness SM fields mix with states of the composite sector of equal quantum numbers under the SM symmetries, see \cite{bib:compositeMFV} for a detailed discussion. All the new states are classified according to representations of the composite sector global symmetry. We will make the minimal assumption that this contains $SU(3)_c\times SU(2)_L\times SU(2)_R \times U(1)_X$. The SM Yukawa couplings are schematically given by
\begin{equation} \label{smyukawas}
	y_{SM}= \sin \phi_L\cdot Y \cdot \sin \phi_R ,
\end{equation}
where $\sin \phi_{L,R}$ are the mixings matrices of left and right chiralities of the SM quarks with the composite states. The coupling $Y$, in general a matrix, has a typical strength  that characterizes the composite sector. For simplicity we will often assume this to be equal to the coupling of spin-1 resonances $g_\rho$ but it should be kept in mind that these are in principle independent parameters.

The standard assumption, naturally realized in Randall-Sundrum scenarios,  is that the degree of compositeness is controlled by the mass of the SM states. Within this logic the light generations are practically elementary and couple only through mixing of the SM gauge fields. This property makes the new states  experimentally well hidden both from direct and indirect searches. It was pointed out however that at least the right-handed chiralities of the light generations could be composite \cite{bib:Atre:2008iu,bib:perez,bib:compositeMFV}. In this case the effects of compositeness are more visible at LHC because the proton constituents are strongly coupled to the composite states. Despite the large degree of compositeness, corrections to precisions observables measured at LEP are small and can be compatible with experimental bounds\footnote{Modified Higgs couplings could also be obtained. See reference \cite{bib:christophe} for the discussion of Higgs precision phenomenology in models with composite right-
handed quarks and reference \cite{bib:Azatov} for related work.}. This perhaps counterintuitive possibility is in fact quite naturally realized if the right-handed quarks couple to singlets of the custodial symmetry. Moreover this possibility is automatic in scenarios that realize the MFV hypothesis \cite{bib:compositeMFV} because a flavor symmetry relates the compositeness of the up quark to the  one of the top that is necessarily large.

Contrary to anarchic scenarios, composite light quarks have striking experimental signatures that could be seen at LHC. Among the new states we will consider the lightest partners of the up and down quarks. For the right-handed quarks we assume that these are singlets of $SU(2)_L\times SU(2)_R$ while left-handed quarks will be associated to bi-doublets. For the up sector we have,
\begin{equation}
\label{eq:rep1}
	L_U = {\bf (2,2)_{\frac 2 3}} = \begin{pmatrix} U & U_{\frac 5 3} \\	D & U_{\frac 2 3} \end{pmatrix} \,,
  \quad \qquad \tilde{U} ={\bf(1,1)_{\frac 2 3}} .
\end{equation}
The full model can be found in the appendix.
\begin{figure}[ht]
	\begin{center}
		\includegraphics[scale=0.25]{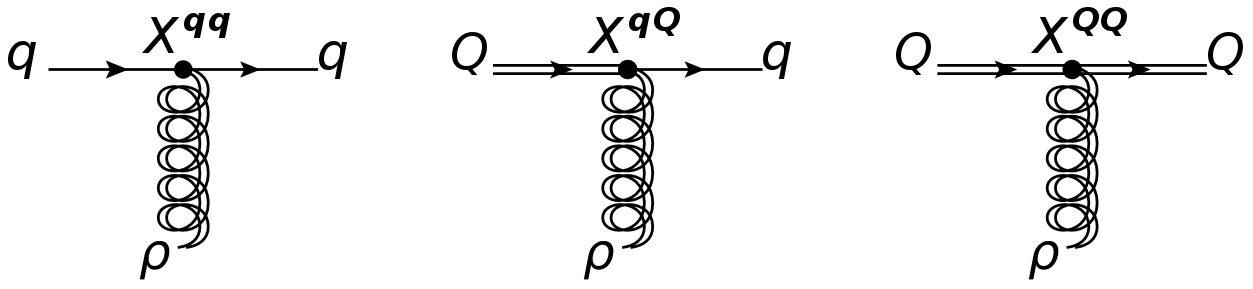}
		\includegraphics[scale=0.25]{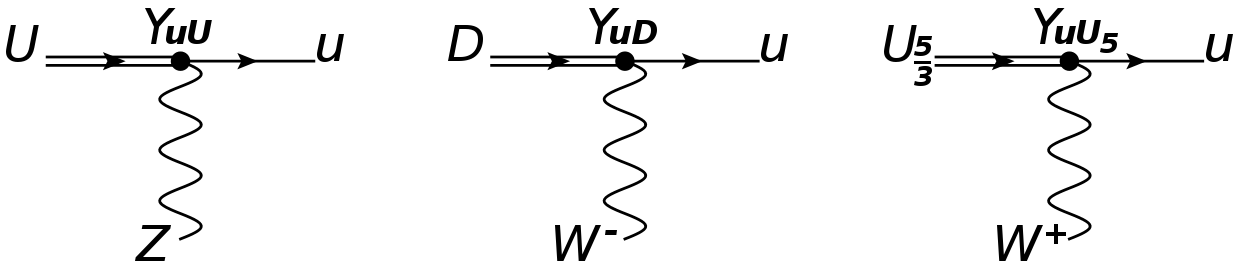}
		\caption{\emph{Above  couplings of the color octet to SM quarks and their heavy partners. Below couplings to electroweak gauge bosons.}}
	\label{fig:rhocouplings}
	\end{center}
\end{figure}
Of the composite spin one states only the gluon partner, a massive color octet vector, will be included. We assume that the color octet couples as a gauge field with strength $g_{\rho}$. Electroweak resonances will not be studied here but we expect the rough features to be similar. The mixing with SM quarks generates the trilinear couplings of the heavy gluon of figure \ref{fig:rhocouplings} with strengths
\begin{align} \label{eq:octetquarkcouplings}
	X_{R}^{qq} & = g_s \left( \sin^2 \phi_{Rq} \cot \theta - \cos^2 \phi_{Rq} \tan \theta \right) \nonumber \\
	X_{R}^{qQ} & = g_s \frac{ \sin \phi_{Rq} \cos \phi_{Rq}}{\sin \theta \cos \theta} \nonumber \\
	X_{R}^{QQ} & = g_s \left( \cos^2 \phi_{Rq} \cot \theta - \sin^2 \phi_{Rq} \tan \theta \right) .
\end{align}
where $\tan\theta =g_s/g_\rho$. We denote by $q$ ($Q$) a light (heavy) quark. Analogous formulas hold for the left-handed chiralities. We will be interested in the situation where the right-handed up and down quarks are significantly composite. Strictly in MFV models $\sin \phi_{Ru} = \sin \phi_{Rt} > \lambda_t / g_\rho$  but this can be relaxed in more general constructions based on $SU(2)$ flavor symmetries \cite{bib:su2}. The SM right-handed quarks  can couple to gluon resonances with a trilinear coupling $q q \rho$ as large as $g_\rho \sin^2 \phi_{Rq}$. Moreover the partners of right-handed quarks can be produced and decay through the heavy-light vertex in figure \ref{fig:rhocouplings}. On the other hand the vertex with left-handed partners is negligible because the compositeness of left-handed light quarks is extremely small.

For electroweak interactions the situation is exactly reverted, see figure \ref{fig:rhocouplings}. In the limit of zero quark masses in the up sector the relevant vertices are
\begin{eqnarray}	\label{eq:electroweak}
	&&  Y_{uD} =  Y_{uU_{\frac{5}{3}}} =\frac g {\sqrt{2}}   \frac {Y_U\,v} {\sqrt{2} m_Q}\sin \phi_{Ru}\nonumber \\
	&& Y_{uU}=- Y_{uU_{\frac 2 3}} =\frac g {2 \cos\theta_W}  \frac {Y_U\,v} {\sqrt{2} m_Q}\sin \phi_{Ru}
\end{eqnarray}
where $v = 246$ GeV and $Y_U$ is the up sector fermionic coupling, see appendix \ref{sec:compositemodel}. These interactions allow to singly produce the partners of left-handed quarks. Higgs interactions are also generated but we will not study them here, for more information see \cite{bib:Atre:2013ap}.

The last important ingredient in our analysis will be the chromomagnetic operator
\begin{equation} 	\label{eq:chromo}
	\mathcal{L}_\mathrm{chromo}^\mathrm{SM} =\kappa \frac {g_s}{m_Q} \,  \bar{U}_L \sigma_{\mu\nu} T^a u_R G_{\mu\nu}^a + \mathrm{h.c.}
\end{equation}
This dimension five operator is relevant in our analysis because it controls the decay of the right-handed partners in the region $m_\rho > m_Q$ where the decay into $Q\to \rho\, q$ is kinematically forbidden. It is generated by loops of the strong sector fields with a size (see appendix \ref{sec:compositemodel})
\begin{equation}
	\kappa \sim \frac {g_\rho^2}{16\pi^2} \frac {m_Q^2}{m_\rho^2} \sin \phi_{uR} .
\end{equation}

Let us briefly comment on the scenario where left-handed quarks are strongly composite. Here precision electroweak tests, in particular modified coupling to the $Z$, strongly disfavours large compositeness. One finds \cite{bib:compositeMFV},
\begin{equation}
	\sin \phi_{Lq} \lsim   \frac {\lambda_t} {2\, g_\rho}  \left(\frac{m_\rho}{\rm 3 \,TeV}\right) .
\end{equation}
Repeating the analysis above implies that cross sections not larger than in the anarchic scenario will be obtained, at least for the scales and couplings that we expect in composite models that address the hierarchy problem. In fact due to the opposite sign of the two contributions in eq. \eqref{eq:octetquarkcouplings} the couplings may even turn out to be smaller. In what follows we will only consider the scenario with composite right-handed quarks.

\subsection{Simulations}
\label{sec:simulations}
In this paper we will study  the phenomenology of the gluon resonance, partners of left-handed quarks $\bf{(2,2)_{2/3}}$ and partners of right-handed quarks, $\bf{1_{2/3}}$ and $\bf{1_{-1/3}}$.  We focus on the first generation partners whose mass is however equal to the one of the top partners under the MFV hypothesis. The searches are very sensitive to the spectrum of the new states. We will mostly work under the assumption that the fermionic scale $m_Q$ is smaller than $m_\rho$. This hypothesis appears to be necessary for the theory to be natural, given that spin one particles lighter than $2$ TeV are disfavoured. On the other hand new vectorial fermions are the most relevant from the naturalness point of view, have weaker direct bounds.

In our simulations we generate event samples with MadGraph5 \cite{bib:mg5}, using a model\footnote{The FeynRules implementation of the right-handed partial compositeness model is available upon request by the authors.} generated with Feynrules 1.6 \cite{bib:feynrules}. The parton level events are passed to Pythia 6.4 \cite{bib:pythia} to simulate the effects of parton showering, and then to Delphes 2.0 \cite{bib:delphes} or ATLFAST \cite{ATLFAST}  for a fast detector simulation. We use the default CMS and ATLAS parameters for Delphes depending on what experimental analysis we are comparing with, and reconstruct jets with the anti-$k_T$ algorithm \cite{bib:antikt} using $0.5$ and $0.7$ for the jet cone radius respectively. These simulated events are then analyzed using the experimental analyses, providing a method to interpret the relevant experimental searches in terms of our model.

\section{Color Octet}
\label{sec:coloroctet}
Among possible spin-1 resonances we will focus on the gluon partner, a color octet with mass $m_\rho$. The experimental searches of dijets and $t\bar{t}$ by CMS and ATLAS imply important bounds on the parameter space of our scenario that we derive in  this section. Constraints on spin one resonances from flavor physics are not necessarily negligible, even if MFV is realized, as certain operators (in particular $(\bar{q}_L y_u y_u^\dagger q_L)^2$)  are generated at tree level \cite{bib:su2}. Nevertheless, these bounds are more model dependent (for example they could be avoided in extensions of MFV) and we will not include them here (see however \cite{bib:Barbieri2012tu}).
\begin{figure}[ht!]
	\begin{center}
		\includegraphics[scale=0.2]{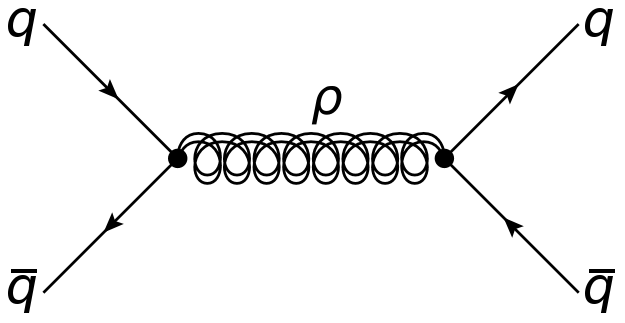} \hspace{1cm}
		\includegraphics[scale=0.2]{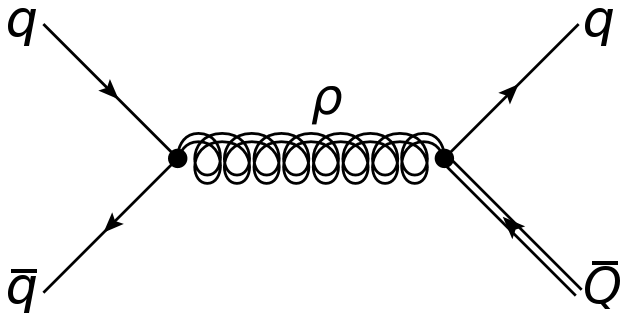} \hspace{1cm}
		\includegraphics[scale=0.2]{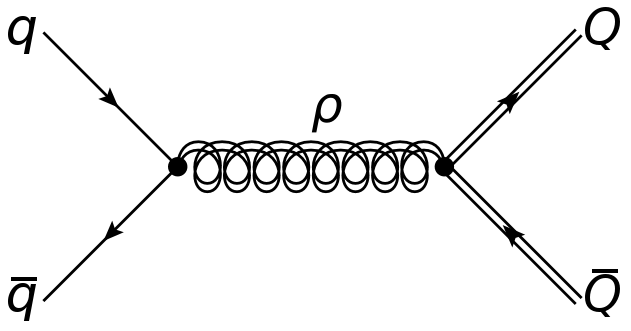}
		\caption{\emph{The heavy color octet is dominantly produced from a quark anti-quark pair and then decays into any kinematically accessible combination of light and heavy quarks.}}
		\label{fig:octetpheno}
	\end{center}
\end{figure}

\subsection{Octet Phenomenology}
\label{sec:octetphenomenology}
The color octet can be produced through the Drell-Yan process $ q \bar{q}\to \rho$ of figure \ref{fig:octetpheno}. Through the coupling with light quarks \eqref{eq:octetquarkcouplings}, it can be copiously produced at LHC if $\sin \phi_{Ru}$ is sufficiently large. No gluon fusion is possible due to gauge invariance.

The decay of the $\rho$ will play an important role in the phenomenology. The decay into SM right-handed quarks is equal for all generations while only the one into $t_L$ is relevant for left-handed quarks. If the heavy fermions are lighter than the color octet the decay into a single heavy and one SM fermion or two heavy fermions (for $2 m_Q<m_\rho$) will be possible. Since the couplings to the composite states are large this can affect strongly the phenomenology.

\begin{figure}[ht!]
	\begin{center}
		\includegraphics[scale=0.5]{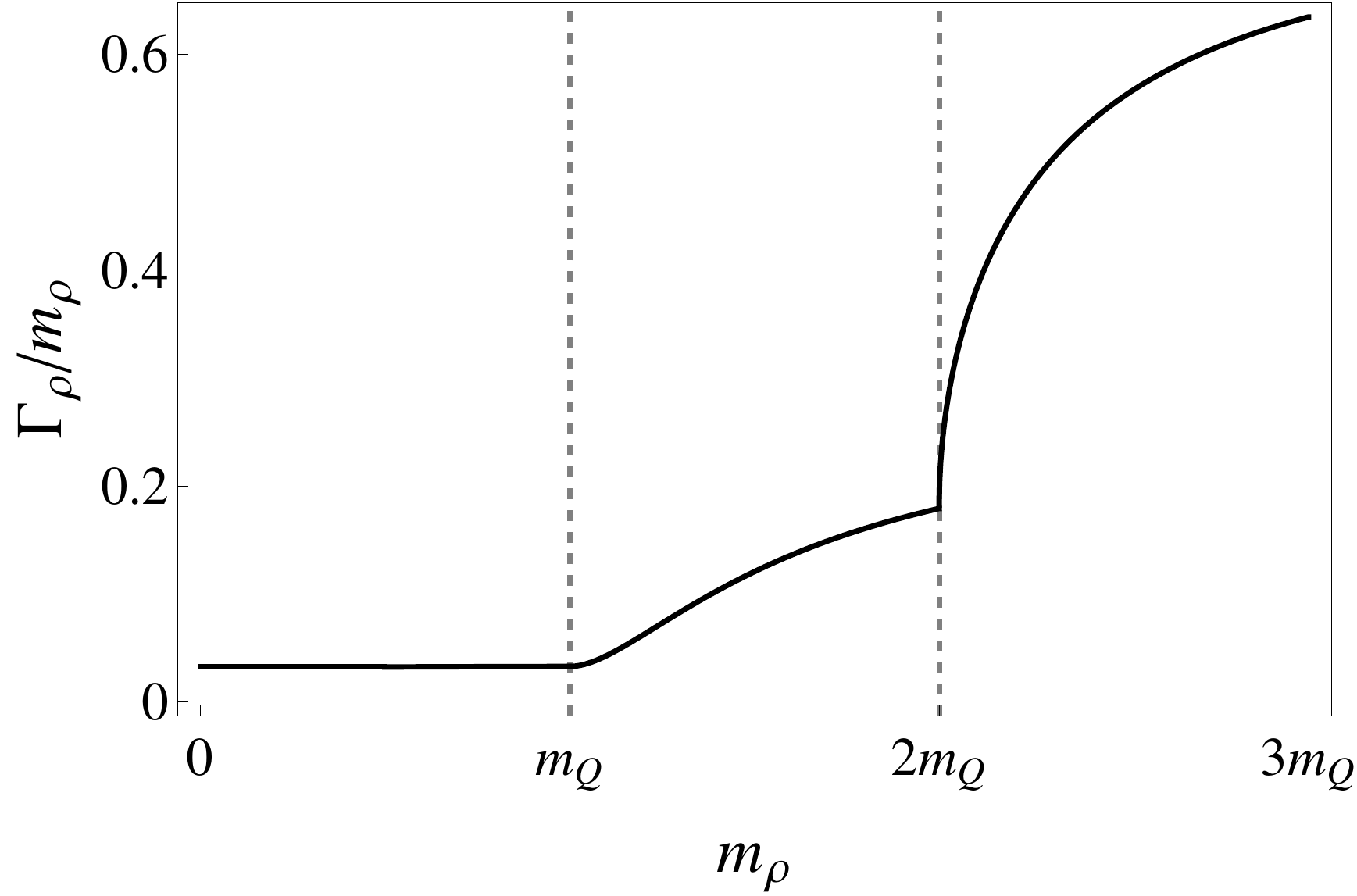}
		\caption{\emph{Typical behavior of the color octet width as a function of the octet mass, for $g_\rho = 3$. The width increases significantly as the decay channels to the quark partners open up, these qualitative features hold independently of the mixings.}}
		\label{fig:octetwidth}
	\end{center}
\end{figure}

The decay modes are displayed in figure \ref{fig:octetpheno}. Analytic formulas for the partial widths read
\begin{align}
	\Gamma (\rho \to q \bar q) & = \frac{\alpha_s}{12} m_\rho \left[ \left( X_L^{qq} \right)^2 + \left( X_R^{qq} \right)^2 \right] \nonumber \\
	\Gamma (\rho \to q \bar Q, \, Q \bar q) & = \frac{\alpha_s}{12} m_\rho \left( 1 - \frac{m_Q^2}{m_\rho^2} \right)  \left( 1 - \frac{m_Q^2}{2m_\rho^2} - \frac{m_Q^4}{2 m_\rho^4} \right) \left[ \left( X_L^{qQ} \right)^2 + \left( X_R^{qQ} \right)^2 \right] \nonumber \\
	\Gamma (\rho \to Q \bar Q) & = \frac{\alpha_s}{12} m_\rho \sqrt{1 - \frac{4 m_Q^2}{m_\rho^2}} \left[ \left( 1 - \frac{m_Q^2}{m_\rho^2} \right)  \left[ \left( X_L^{QQ} \right)^2 + \left( X_R^{QQ} \right)^2 \right] + 6 \frac{m_Q^2}{m_\rho^2} X_L^{QQ} X_R^{QQ} \right] .
\end{align}
in the limit $m_q \ll m_Q$. $X_{L/R}$'s are the couplings as defined in equation \eqref{eq:octetquarkcouplings}. As shown  in figure \ref{fig:octetwidth} the width of the color octet changes drastically when the decay modes to one or two heavy fermions open up. In the last case the resonance is very broad.

\subsection{Compositeness Bounds}
\label{sec:compositenesbounds}

We start our analysis considering compositeness bounds reported by ATLAS and CMS, see also \cite{bib:compositeMFV,bib:Domenech:2012ai}. In the large $m_\rho$ limit we can integrate out the color octet and replace it with an effective four fermion operator. Such an operator produces dijets with an angular distribution different from the QCD that allows to distinguish it from the background. The experiments in particular place a bound on the effective operators with light quarks,
\begin{equation}
	c_{LL} \, (\bar{q}_L \gamma^\mu q_L)^2 + c_{RR} \, (\bar{q}_R \gamma^\mu q_R)^2 + 2 c_{LR} \, (\bar{q}_L \gamma^\mu q_L) (\bar{q}_R \gamma_\mu q_R) .
\end{equation}
that can be recast in  our scenario. Recent experimental results on the angular distributions of dijet final states by both ATLAS \cite{bib:atl2jang} and CMS \cite{bib:cms2jang} imply
\begin{align}
	c_{LL,RR}^{(+)} \lsim 0.10 \, \mathrm{TeV}^{-2} & \quad \mathrm{ATLAS} \nonumber \\
	c_{LL,RR}^{(-)} \lsim 0.06 \, \mathrm{TeV}^{-2} & \quad \mathrm{CMS} .
\end{align}
The $\pm$ superscript refers to the sign of the coefficient, and the ATLAS analysis only considers the case of destructive interference. CMS provides an exclusion for both signs of the coefficient and the most constraining one is used. Note that the operators with heavy quarks (such as $u\bar{u} c\bar{c}$) are expected to be less relevant at LHC since dijet production requires a quark-antiquark initial state in that case, a process suppressed by the protons PDFs. Integrating out the heavy color octet one generates the four fermion operator \cite{bib:compositeMFV}
\begin{equation} \label{eq:octetoperator}
	\frac{g_\rho^2}{6 m_\rho^2} \sin^4 \phi_{L,Rq} \left( \bar{q}_{L,Rq} \gamma^\mu q_{L,Rq} \right)^2 .
\end{equation}
Using the strongest bound reported by CMS and the coefficient in equation \eqref{eq:octetoperator} we derive
\begin{equation}
	\sin^2 \phi_{Ru} \lsim  \frac{0.6}{g_\rho} \left( \frac{m_\rho}{\mathrm{TeV}} \right) .
\end{equation}
Both constraints by ATLAS and CMS are displayed in figure \ref{fig:dijetbounds}. The compositeness of right-handed down quarks is slightly less constrained due to the predominance of up quarks in the proton.

Note that the sign of coefficient obtained integrating out heavy vectors (a similar conclusion holds for scalars) is fixed and corresponds to the most constrained sign in the CMS analysis. Hence, it is useful that the experiments report the bound for both signs of the operator.

\subsection{Resonance Searches}
\label{sec:dijetsearches}
\begin{figure}[ht]
	\begin{center}
		\includegraphics[scale=0.46]{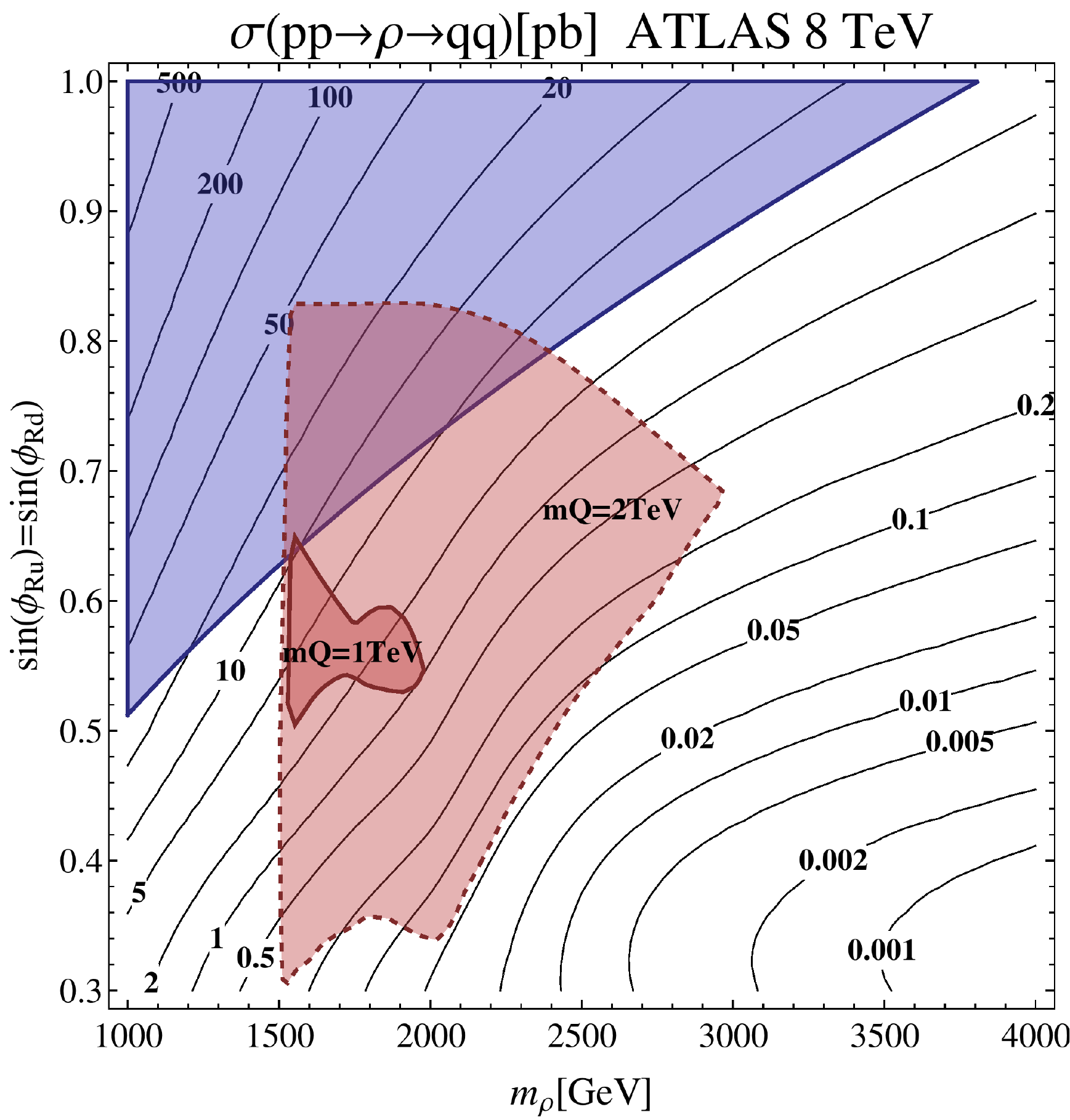} \hspace{.2cm}
		\includegraphics[scale=0.46]{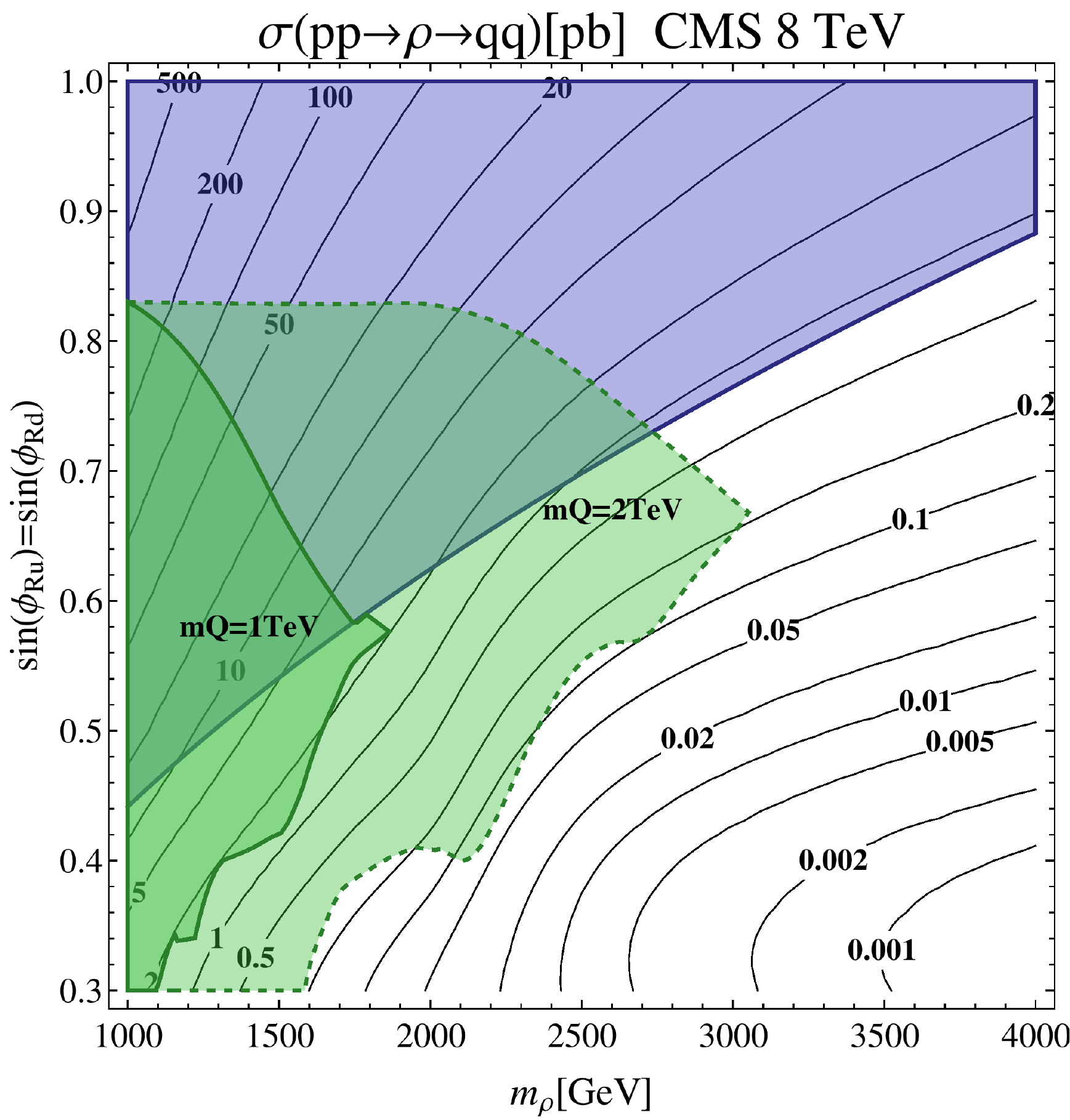}
		\caption{\emph{Exclusion plot for a color octet with $g_\rho=3$. In blue, region excluded by compositeness bounds.
		In red (ATLAS) and green (CMS) exclusion from direct production. The different regions correspond to $95$\% confidence level exclusion for two hypothetical scenarios where the quark partners are light ($m_Q = 1$ TeV, solid contour) or heavy ($m_Q = 2$ TeV, dashed contour).}}
		\label{fig:dijetbounds}
	\end{center}
\end{figure} \noindent

If the resonance is sufficiently light it can be produced in $pp$ collisions and then decay into jets. The natural search strategy is to look for a bump in the invariant mass distribution of dijets. We emphasize that this search is of a very different nature compared to compositeness bounds that rely on the angular distribution of dijets to distinguish new physics effects from the enormous QCD background. While the effective operator bound is limited by the energy of the accelerator, the on-shell production, when kinematically accessible, is limited by statistics.

We use the most recent analyses of ATLAS \cite{bib:atl2jets} and CMS \cite{bib:cms2jets,bib:cms2jetsupdate} based on  $8$ TeV data. The experiments provide a limit on $\sigma(pp \to \rho) \times BR(\rho\to qq)$ of resonances coupled to light quarks that can be applied to our scenario. We follow the procedure given by ATLAS for a Gaussian resonance with a particular width ranging from zero to 15\%. Roughly the same strategy is applied to the CMS search, which provides limits on $\sigma \times BR \times \epsilon$. Our bounds are conservative as we explicitly take into account the width of the resonance. Both ATLAS and CMS perform a search for a relatively narrow resonance through a bump hunter algorithm. If the width of the the resonance exceeds a certain threshold the bump hunter search is invalidated and hence we discard the limit whenever the color octet width is above $15$\%. In the $m_\rho$ and mixing angle plane, a grid of points is generated for which $\sigma \times Br$ are computed, and the 
efficiencies of the experimental cuts are analyzed. The results are compared to the experimental limit, that only depends on the resonance mass, and are then interpolated to form exclusion regions.

The limits for a resonance with $g_\rho=3$ are presented in figure \ref{fig:dijetbounds}. The blue region corresponds to the bound on the effective four fermion operators discussed in the previous section. The exclusion due to the on-shell production is given by the red and green regions. This exclusion limit depends strongly on the fermionic spectrum, because of two reasons. One is the increase in the width of the resonance possibly invalidating the search. This becomes particularly relevant when the decay into two heavy partners is kinematically accessible, see figure \ref{fig:octetwidth}. Moreover when other channels open up the signal strength is reduced since only the decay into SM quarks will generate a bump in the invariant mass distribution of the two leading jets. For this reason the region with $m_Q=1$ TeV is weakly constrained\footnote{We do not include here the partners of left-handed down quarks that would further increase the width if the decay into two heavy quarks is kinematically accessible.}
. Note also that model independently the region of high compositeness is not constrained because the width is in this case always too large.

\subsection{\texorpdfstring{$t \bar{t}$}{ttbar} Searches}
\label{sec:ttbarsearches}
\begin{figure}[ht]
	\begin{center}
		\includegraphics[scale=0.46]{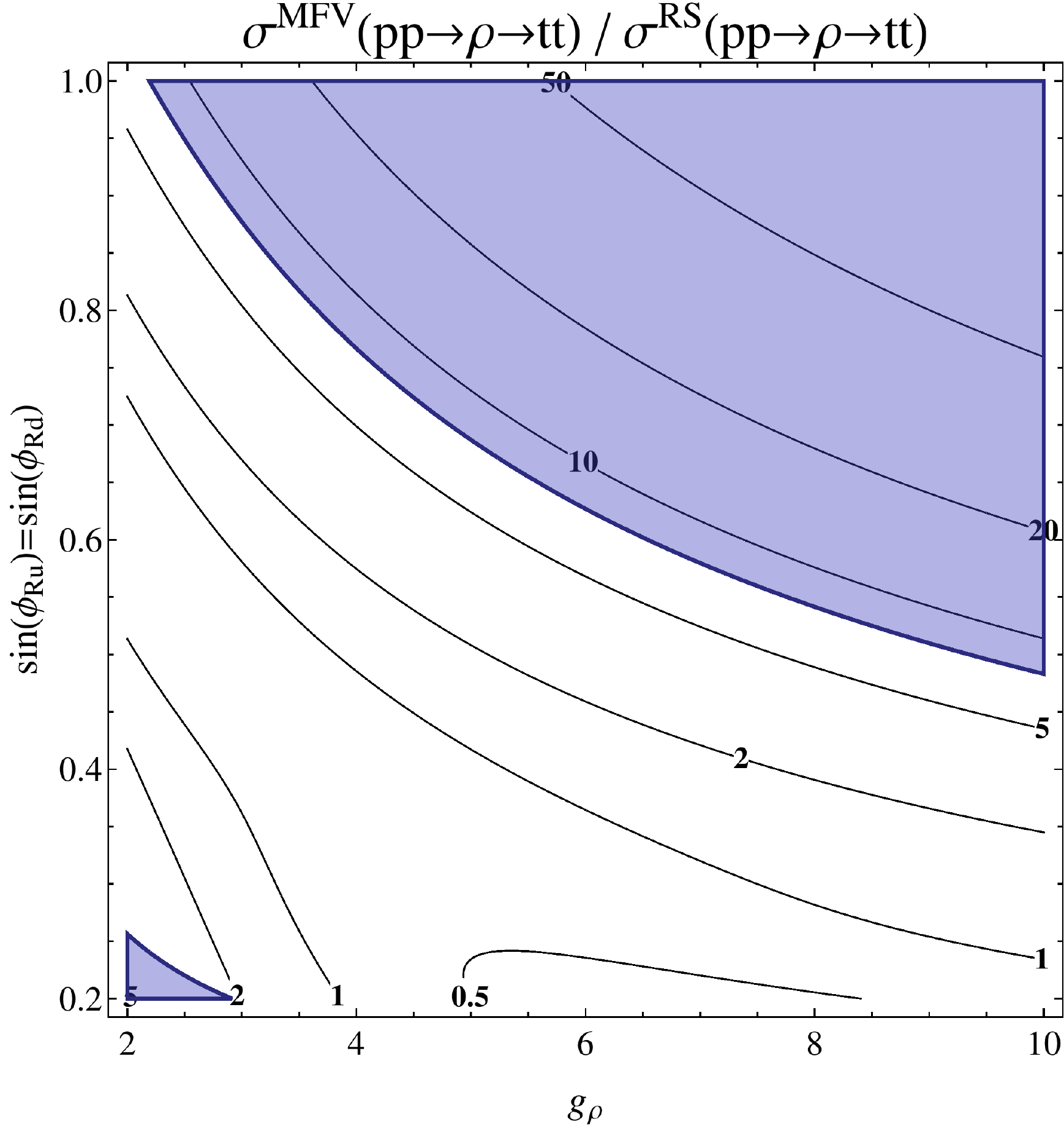}
		\caption{\emph{Ratio of $\sigma(pp \to \rho \to t {\bar t})$ in MFV models compared to the experimental benchmark Randall-Sundrum model. The blue region corresponds to a width greater than
		$0.2 \; m_\rho$ where the experimental bounds are not be applicable. In this comparison the assumption is made that the decay to heavy fermions is kinematically forbidden.}}
		\label{fig:ttbarmfv}
	\end{center}
\end{figure} \noindent

In anarchic scenarios gluon resonances are strongly coupled to the third generation and decay mostly into top quarks. For example in Randall-Sundrum scenarios as considered in reference \cite{bib:gluonresonance} one finds that the branching of heavy gluons into top right is almost $100$\%. To connect with our parametrization this model roughly corresponds to $g_\rho = 5$, $\sin \phi_{Rq} \approx 0$ for the light quarks and $\sin \phi_{Rt}=1$ for the top quark. A strong bound on gluon resonances is obtained through searches of resonances that decay into $t \bar{t}$ pairs. Exclusion limits for this benchmark point have been reported in the searches from ATLAS \cite{bib:atlttbar} and CMS \cite{bib:cmsttbar}. In the case of the Randall-Sundrum benchmark the heavy gluon resonance
is excluded below $1.5$ TeV at $95$\% confidence level.

In models that realize MFV, or more generally models with composite light quarks, the situation is different both for the production and decay of the heavy gluon, and one may obtain an even stronger bound. In these models the decay into third generation is typically not dominant. This depletion of the signal is however easily compensated by the increased production cross section. To get an idea of the bounds in this case, we can estimate $\sigma(pp \to \rho) \times BR(\rho \to t \bar{t})$ by rescaling the couplings of the anarchic scenario\footnote{For simplicity we assume equal compositeness of up and down type right-handed quarks. The result is then approximately independent of PDFs.}. The numerical result is presented in figure \ref{fig:ttbarmfv}. We see that the cross section in $t\bar{t}$ is typically larger than in anarchic scenarios. As a consequence slightly stronger bound will apply.

\begin{figure}[ht]
	\begin{center}
		\includegraphics[scale=0.46]{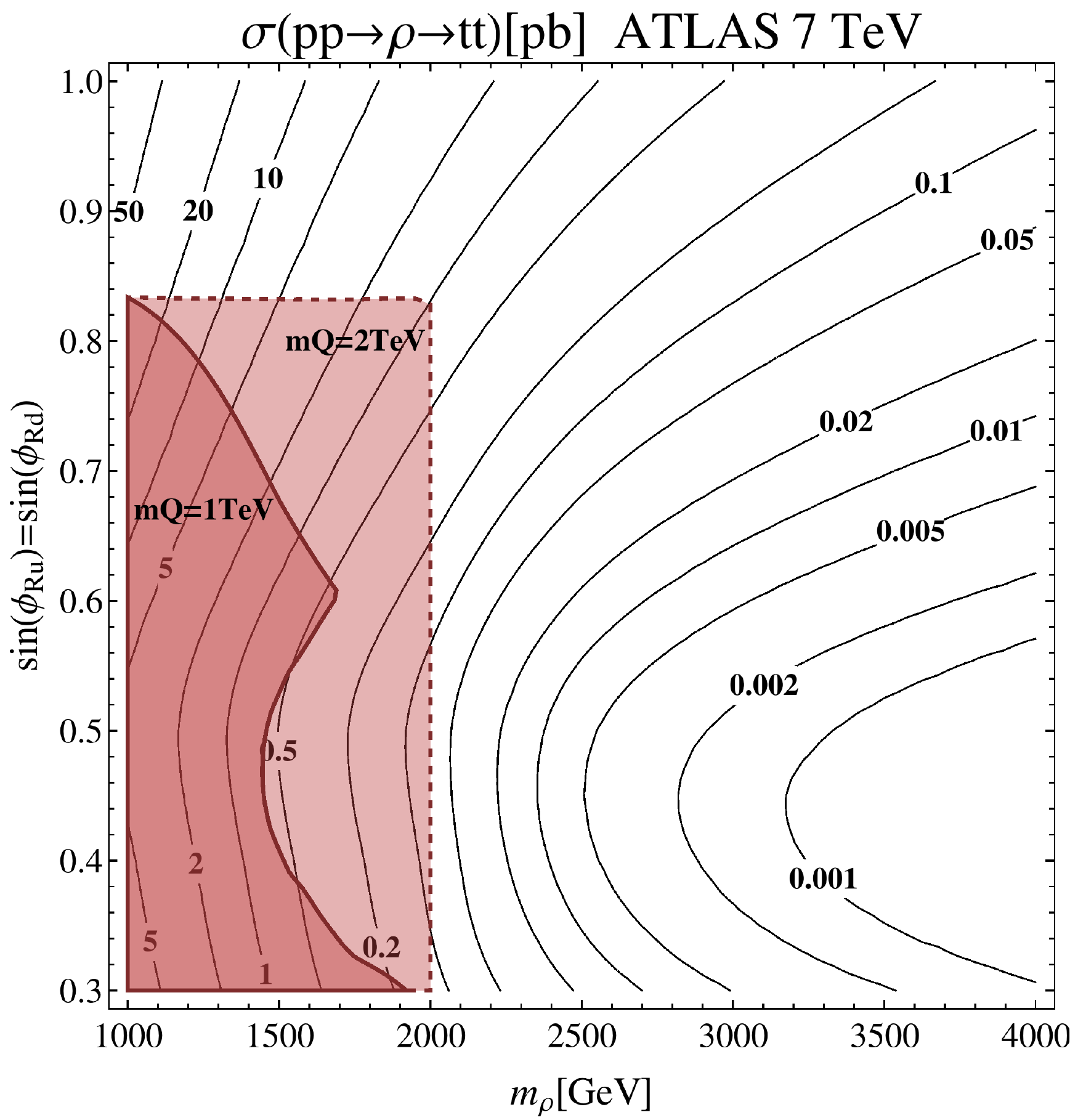} \hspace{.2cm}
		\includegraphics[scale=0.46]{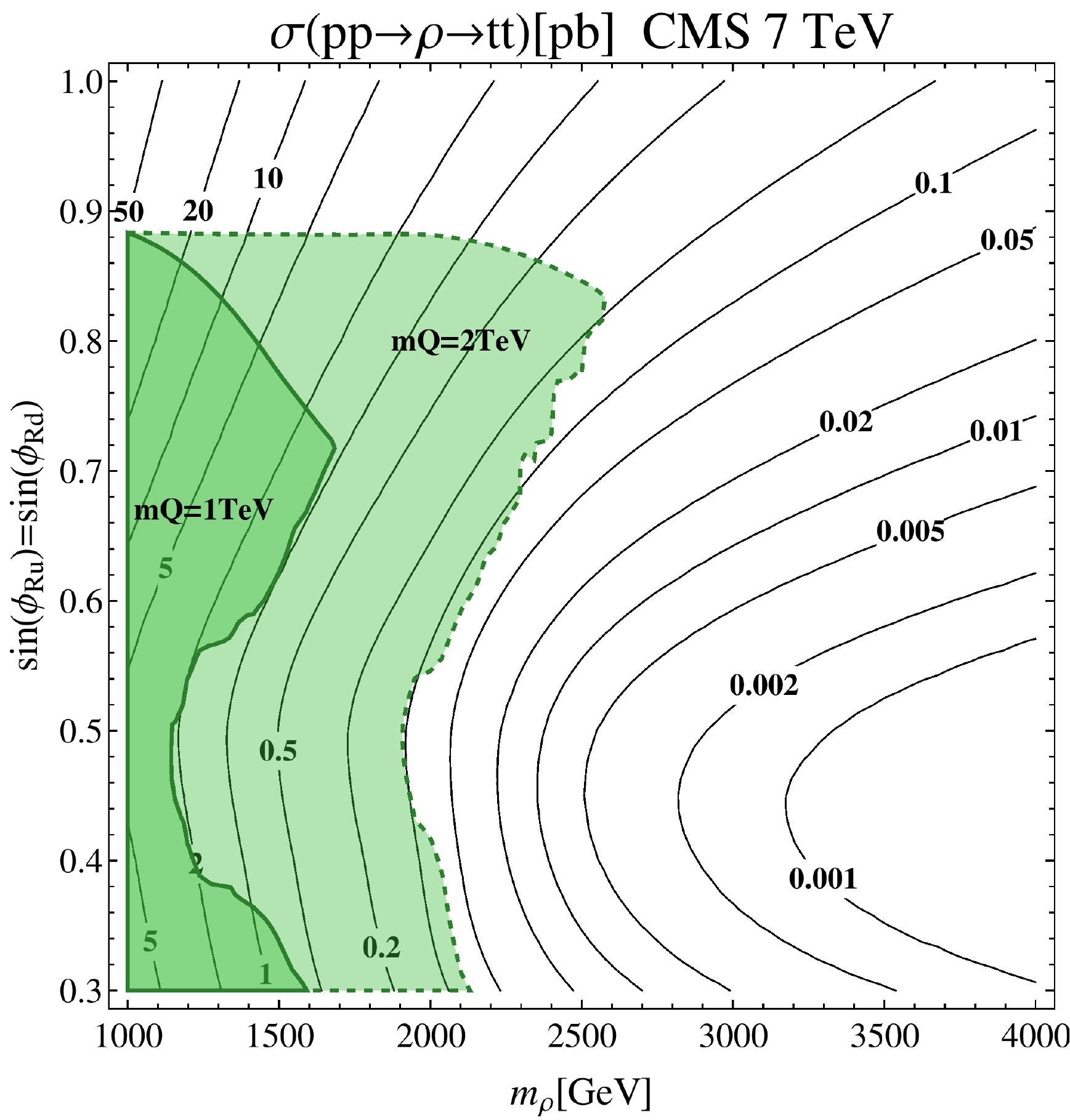}
		\caption{\emph{Constraints from $t \bar{t}$ searches by ATLAS (red) and  CMS (green). The exclusion limits correspond to $95$\% confidence level for two hypothetical scenarios where the quark partners are light ($m_Q = 1$ TeV, solid contour) and where they are heavy ($m_Q = 2$ TeV, dashed contour).}}
		\label{fig:ttbarbounds}
	\end{center}
\end{figure}

One caveat must be considered: similarly to the dijet searches, the experimental bound is obtained by looking for bumps in the invariant mass spectrum of $t\bar{t}$ pairs. This procedure depends on the width of the resonance and becomes inefficient for large widths. In anarchic scenarios the resonances are relatively broad. In the example of \cite{bib:gluonresonance} the width is below $20$\% of the mass. The width can be larger in the MFV scenario due to multiplicity factors and the decay to heavy quark partners, as explained in the section above. The blue region in figure \ref{fig:ttbarmfv} corresponds to a width greater than $0.2 \; m_\rho$. In this region the experimental bound must be reconsidered. This region is however excluded by compositeness bounds discussed in the previous section. To compare the limits with the dijet searches also an exclusion plot in the $(m_\rho , \sin \phi)$ plane is provided in figure \ref{fig:ttbarbounds}. These exclusion plots have been obtained in a
similar fashion as for the dijet limits from the previous section, including a careful treatment of the width of the heavy partners possibly invalidating the $t \bar{t}$ search.

We should mention that in extensions of the MFV scenario based on $SU(2)$ rather than $SU(3)$ flavor symmetries the compositeness of the third generation can be different from the first two \cite{bib:su2}. Those scenarios are attractive phenomenologically as the light generations can be mostly elementary, avoiding compositeness bounds but with the same virtues as MFV for what concerns flavor. In this case the phenomenology of heavy gluons will be similar to anarchic scenarios.

\subsection{Combined Bounds}
\label{sec:combinedbounds}
Summarizing the direct limits on the color octet depend heavily on the fermionic spectrum. We differentiate two scenarios, one with light fermionic partners, $1\,{\rm TeV}<m_Q< 2\,{\rm TeV}$ and one with heavier partners $m_Q> 2\,{\rm TeV}$. In the first we find,
\begin{equation}
	m_\rho > 1500 \textrm{ GeV}
\end{equation}
at $95$\% confidence level. For the heavy scenario we find the constraint
\begin{equation}
	m_\rho > 2000 \textrm{ GeV}
\end{equation}
that is slightly stronger than the bound in anarchic scenarios. This constraint holds for all values of the mixings $\sin \phi_{Rq}$ and tighter bounds on the octet mass are obtained for specific mixings. More stringent bounds can be inferred from from flavor physics and precision tests but these rely on extra assumptions on the structure of the theory and do not directly test the hypothesis of large compositeness of the first and second generation.

\section{Bounds on Left-Handed Quark Partners}
\label{sec:lefthandedpartners}

We start our study of fermionic partners with the left-handed sector, focusing in particular on the $\bf{(2,2)_{2/3}}$ colored fermions. These states can of course be pair produced through strong interactions, see \cite{bib:doubleproduction} for a study in anarchic scenarios. As we will see in the next section the exclusion on top partners of reference \cite{bib:atlas23} can be translated in MFV scenarios into an exclusion of these states around 600 GeV. Here we derive the bound obtained from single production through the electroweak vertices of figure \ref{eq:octetquarkcouplings}. Using the results in \cite{bib:ATLAS:2012apa} we derive a bound significantly stronger than the one on top partners.

The single production of left-handed partners is dominated by $t$-channel exchange of electroweak gauge bosons producing a forward jet ($p_T\sim m_W$) and a heavy quark. This decays mainly through weak interactions into jets and $W$, $Z$ or Higgs\footnote{We do not include the decay into Higgs in our analysis. This was recently studied in \cite{bib:Atre:2013ap}.}. The jet has the same flavor of the mother particle so that only light quark jets are obtained in the final state.

\begin{figure}[ht]
	\begin{center}
		\includegraphics[scale=1]{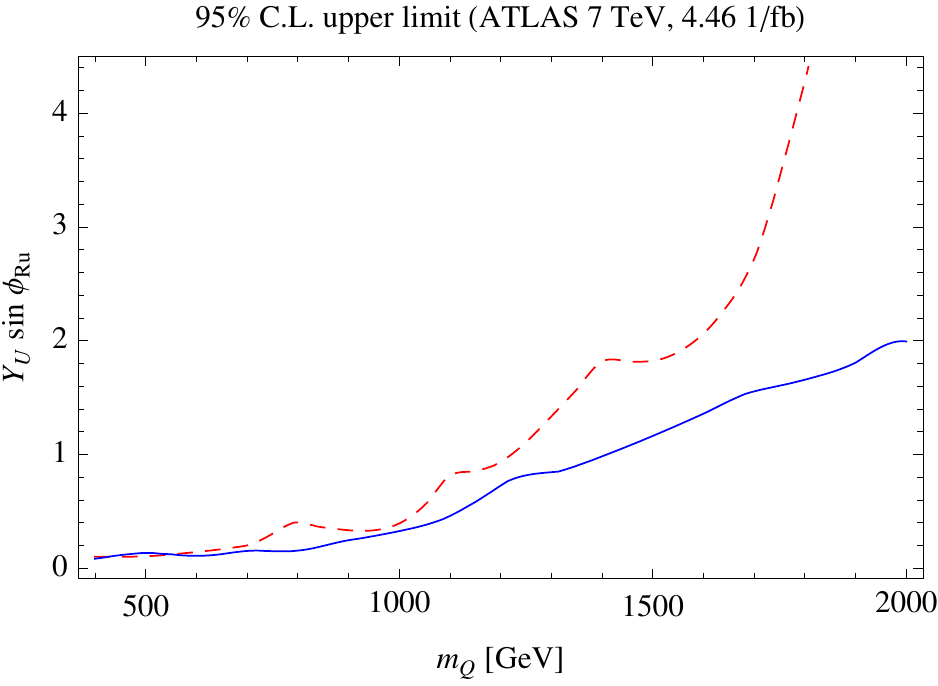}
		\caption{\emph{Exclusion of  left-handed partners of the up quark by ATLAS \cite{bib:atlas23}.  The red dashed (blue solid) line shows the 95 \% C.L. observed upper limit on $Y_U \sin \phi_{Ru}$ obtained from the search of charge 2/3 (5/3) heavy quarks. The regions above the lines are excluded.}}
		\label{fig:crossx}
	\end{center}
\end{figure}

In the  anarchic scenarios only third generation quarks can be produced in this way because the coupling is proportional to the degree of compositeness that is significant only for the third generation. To produce third generation partners one needs to scatter a $W$ or $Z$ boson and a top, the latter originating from the splitting of a gluon. This will be perhaps the most promising channel for the production of heavy fermions at LHC14 \cite{bib:singleproduction} but in the present run suppression from PDFs and low luminosity is too severe for this process to be dominant. This is different with composite light quarks since the heavy partners can be directly produced with the proton constituents. In this case one can produce the left-handed partners through diagram b) in figure \ref{fig:fermionprod} with access to the valence quarks of the proton. A related aspect is that the width of the resonance is larger than in anarchic scenarios.

The search of vector like quarks coupled to the first generation was  performed by the ATLAS collaboration based on \cite{bib:azuelos} (see also \cite{bib:Adam,bib:Carmona2012jk}) and can be applied to our scenario. We use the most recent results in \cite{bib:ATLAS:2012apa} obtained with 5 fb$^{-1}$  luminosity and 8 TeV energy. The search constrains directly the combination $Y_U \sin \phi_{Ru}$. The derived exclusion is shown in figure \ref{fig:crossx} for the charge 2/3 and exotic charge 5/3 states, the latter being the strongest. Recall that in MFV scenarios there is a constraint,
\begin{equation}
	Y_U \sin \phi_{Ru} \gsim 1
\end{equation}
necessary to reproduce the top mass. From this it follows that the left-handed partners are often excluded up to 2 TeV and always below 1.5 TeV. This can only be avoided in extensions of MFV where the third generations can be split \cite{bib:su2}.

We emphasize that this is an extremely strong bound that pushes the model into fine tuning territory. In view of the recent discovery of a $125$ GeV resonance \cite{Higgs-discovery} some of the fermions associated to the top should be light if the theory shall remain natural. Recent analyses have shown that the lightest top partner should be typically below $1$ TeV in a natural theory \cite{bib:lighthiggs}. In MFV scenarios the mass of the top partners is the same as the one of the light generations, up to mixing effects. Hence, we can translate the bound on the light generations into a bound on the top partners.

\section{Bounds on Right-Handed Quark Partners}
\label{sec:righthandedpartners}
The phenomenology of partners of right-handed quarks is entirely different as they cannot be singly produced by electroweak interactions and they mostly decay into two or three jets leading to multi-jet final states. The majority of multi-jet searches at LHC, being motivated by supersymmetry, assumes a large missing energy typically of the order of few hundreds GeV or more. In our scenario, the missing energy in the event is a consequence of jet calibration accumulated by all jets, typically below 50 GeV. Therefore, we do not expect vanilla supersymmetric searches to play a role in constraining the parameter space of right-handed compositeness. Analysis of the relevant ATLAS and CMS searches will be done in the next two sections, separated into single production (through heavy resonances) and double production (both through QCD and heavy resonances). Dedicated searches that could improve the experimental reach will be discussed in the section 6. Before analyzing the different searches at the LHC we first 
review production modes and decay channels in detail.
\begin{figure}[ht]
	\begin{center}
		a) \includegraphics[scale=0.2]{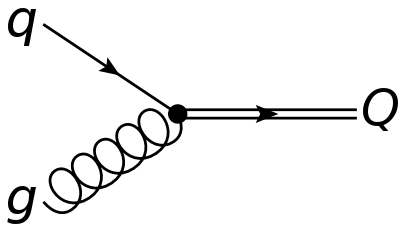} \hspace{.2cm}
		b) \includegraphics[scale=0.2]{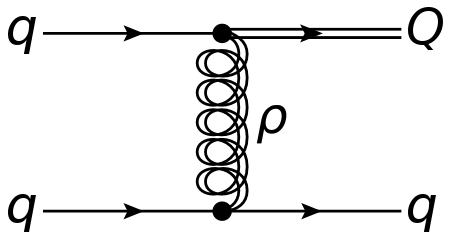} \hspace{.2cm}
		c) \includegraphics[scale=0.2]{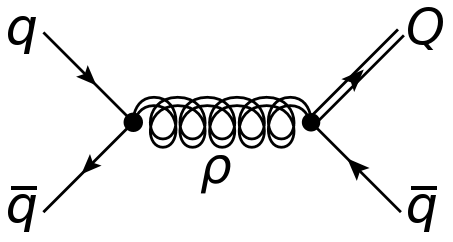} \hspace{.2cm}
		d) \includegraphics[scale=0.2]{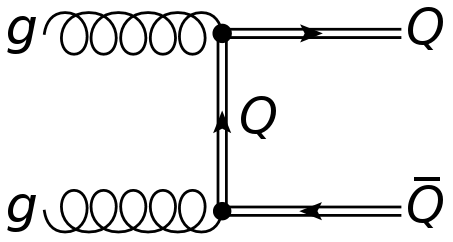}
		\\ \vspace{0.5cm}
		e) \includegraphics[scale=0.2]{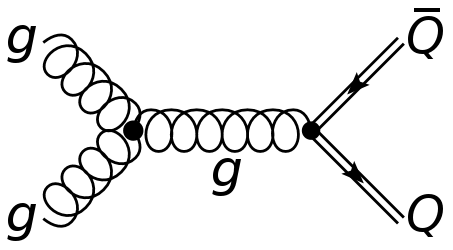} \hspace{.2cm}
		f) \includegraphics[scale=0.2]{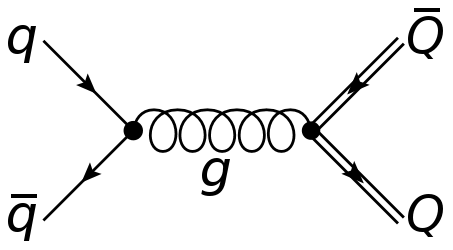} \hspace{.2cm}
		g) \includegraphics[scale=0.2]{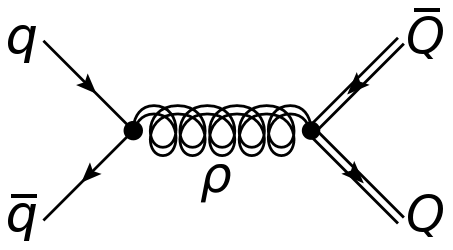} \hspace{.2cm}
		h) \includegraphics[scale=0.2]{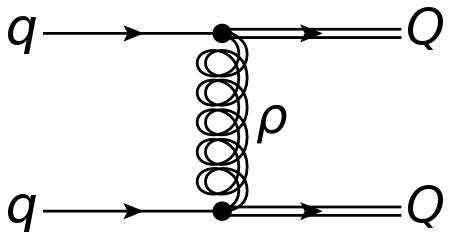}
		\caption{\emph{Fermion production modes: $a)$ chromomagnetic $s$-channel, $b)\,$-$\,c)$ single production and $d)\,$-$\,h)$ double production.}}
		\label{fig:fermionprod}
	\end{center}
\end{figure}

\begin{figure}[ht]
	\includegraphics[scale=0.46]{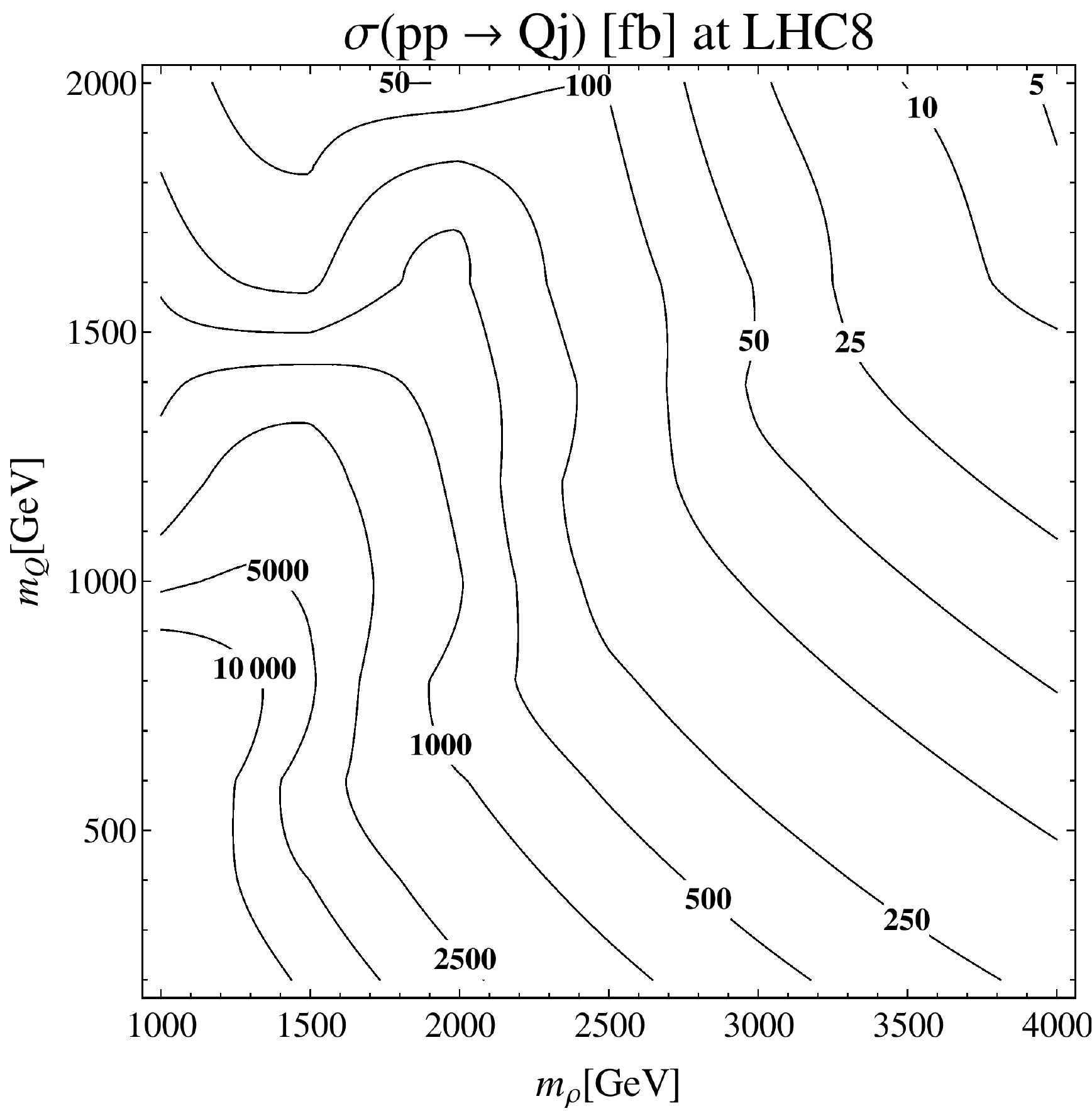} \hspace{.5cm}
	\includegraphics[scale=0.46]{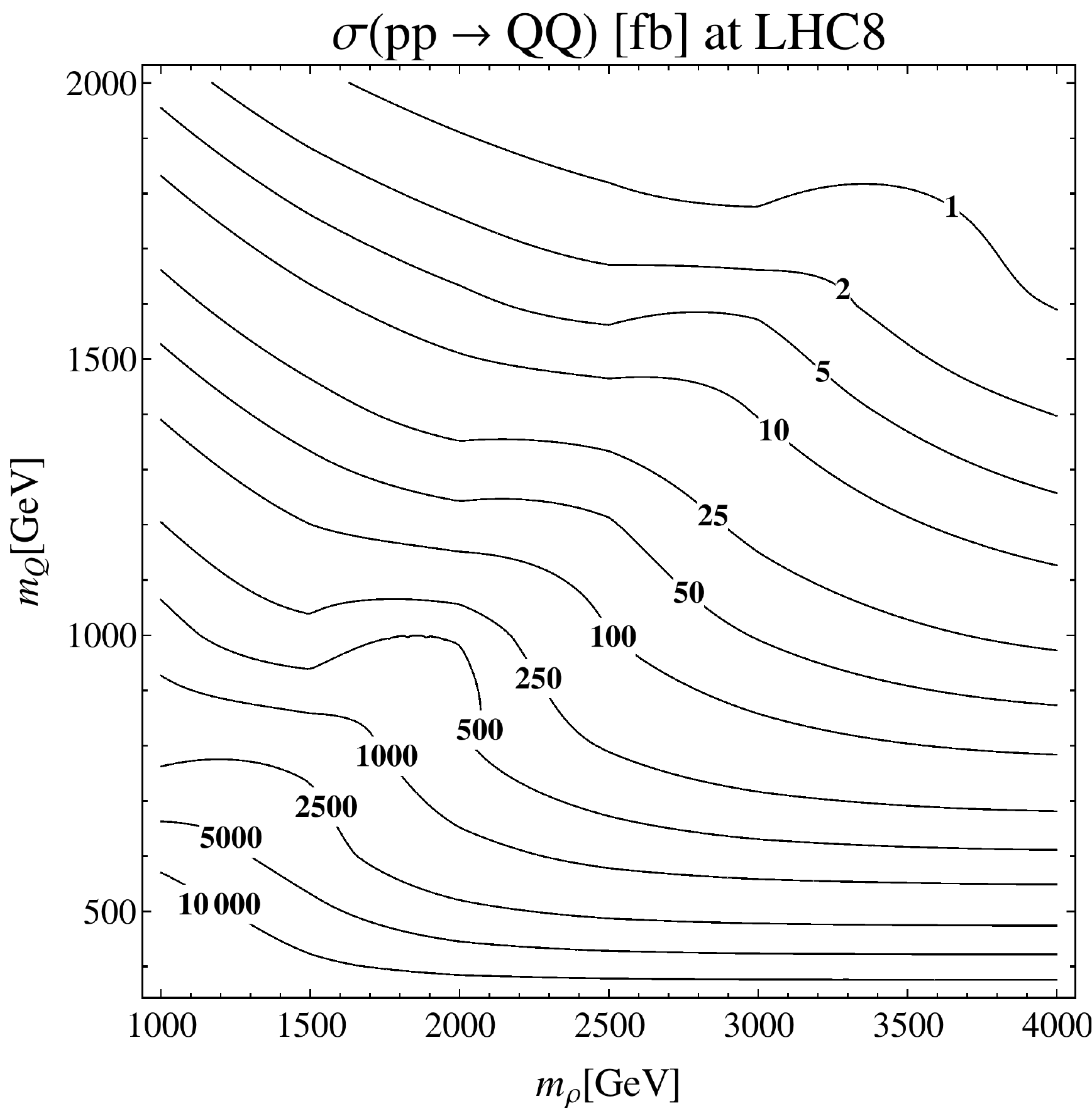}
	\caption{\emph{On the left total cross section of associated production of a heavy quark partner at LHC8 obtained with MadGraph \cite{bib:mg5} for $g_\rho =3$, $\sin \phi_{Ru} = \sin \phi_{Ru} = 0.6 $. On the right double production of right-handed quarks partners ($t_R$ excluded) through QCD
	and heavy gluon exchange.}}
	\label{fig:fermionprodcrossx}
\end{figure}

\paragraph{Production Modes:}
The heavy fermions associated to the first generation can be singly produced in association with a quark via a $t$-channel exchange of the color octet\footnote{Single production via the chromomagnetic interaction \eqref{eq:chromo} will be subdominant under the assumption that the coefficient is loop suppressed.}. Double production of heavy fermions proceeds through $s$-channel gluon or color octet exchange or a $t$-channel color octet or heavy fermion. Both the production modes with either an $s$-channel or a $t$-channel color octet dominate. The various production modes are depicted in figure \ref{fig:fermionprod}. The relevant production modes can be summarized in associate single production and double production. For these modes the production cross section as a function of the color octet mass and the heavy quark mass is given in figure \ref{fig:fermionprodcrossx}.

\begin{figure}[ht]
	\begin{center}
		\includegraphics[scale=0.25]{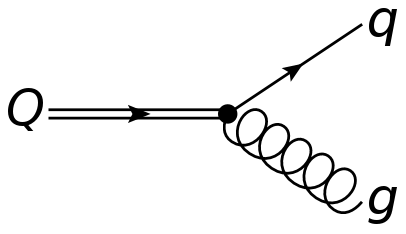} \hspace{1cm}
		\includegraphics[scale=0.25]{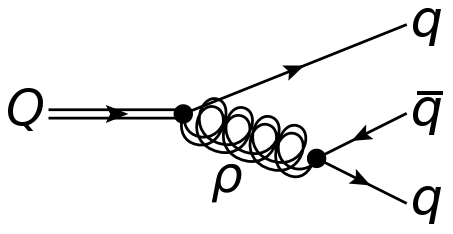} \hspace{1cm}
		\includegraphics[scale=0.25]{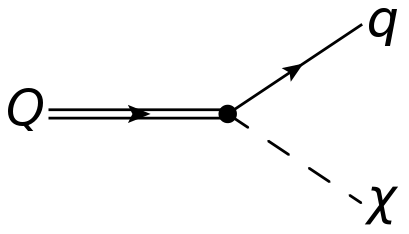}
		\caption{\emph{Fermion decay channels: two body decay via the chromomagnetic operators, three body decay via an off-shell color octet and electroweak two body decay.}}
		\label{fig:fermiondecay}
	\end{center}
\end{figure}

\paragraph{Decay Channels:}
The heavy partners of SM right-handed quarks are singlets of $SO(4)$. Due to this fact they decay almost entirely into jets. The different decay channels are displayed in figure \ref{fig:fermiondecay}. The chromomagnetic interaction induces a decay to a gluon and a quark and generates a width
\begin{equation}
	\Gamma_\mathrm{chromo} (Q \to q g) = \frac{4}{3} \alpha_s \kappa^2 \sin^2 \phi_{Ru} \, \frac{1}{m_Q^5} \left| m_Q^2 - m_q^2 \right|^3 ,
\end{equation}
and the same for the down type quarks with $\phi_{Ru} \to \phi_{Rd}$. This decay is induced at one loop and is typically very small, competing with three body decay mediated by an off-shell $\rho$. An analytical expression for the three body decay is quite lengthy and therefore we only give the limiting behavior (with all light quark masses set to zero $m_q = m_{q'} = 0$ and narrow width approximation for the $\rho$: $\Gamma_\rho \ll m_\rho$)
\begin{equation}
	\Gamma_\textrm{3-body}^\rho (Q \to q q' \bar q') = \left\{
	\begin{aligned}
   & \frac{\alpha_s^2}{72 \, \pi} \left[ \left( X_L^{qQ} \right)^2 + \left( X_R^{qQ} \right)^2 \right] \sum_{q'} \left[ \left( X_L^{q'q'} \right)^2 + \left( X_R^{q'q'} \right)^2 \right] & \\
   & \times \left[ \frac{6 m_\rho^4 - 3 m_Q^2 m_\rho^2 - m_Q^4}{m_Q m_\rho^2} + \frac{m_\rho^2 ( m_\rho^2 - m_Q^2 )}{m_Q^3} \log \frac{m_\rho^2 - m_Q^2}{m_\rho^2} \right] & \mathrm{if} \; m_Q < m_\rho \\
   & \frac{\alpha_s}{6} \left( \frac{m_Q^6 - 3 m_Q^2 m_\rho^4 + 2 m_\rho^6}{m_Q^3 m_\rho^2} \right) \left[ \left( X_L^{qQ} \right)^2 + \left( X_R^{qQ} \right)^2 \right]  & \mathrm{if} \; m_Q \gg m_\rho
  \end{aligned} \right.
\end{equation}
The full analytic expression including the width of the heavy color octet has been used for the analyses. This decay suffers from the octet being off-shell and phase space suppression. Finally a decay to SM quarks plus a longitudinal W, Z or Higgs \cite{bib:2sitescontino} is possible
\begin{equation}
	\Gamma_\mathrm{2-body}^\mathrm{EW} (Q \to q H ) \approx \frac{1}{4 \pi} \frac{m_q^2}{v^2} \frac{\cos^2 \phi_{Ru}}{\sin^2 \phi_{Ru}} m_Q .
\end{equation}

In the MFV scenario the electroweak two body decay is entirely negligible for the first generation as it is suppressed by the light quark mass over the vacuum expectation value. It can also be subleading for the second while it is certainly dominant for the third generation. Note that this conclusion does not hold in the anarchic scenario, in that case $\sin \phi_{Ru}$ is smaller and the decay through electroweak interactions dominates producing $W, Z, h + jets$ final states.

To avoid model dependence in what follows we only focus on light generation partners. For single production the situation effectively reduces to this while for double production this is a conservative assumption and larger cross sections can often be obtained due to the flavor multiplicity. Because of this our conclusions can be considered conservative.

\begin{figure}[ht]
	\begin{center}
		\includegraphics[scale=0.46]{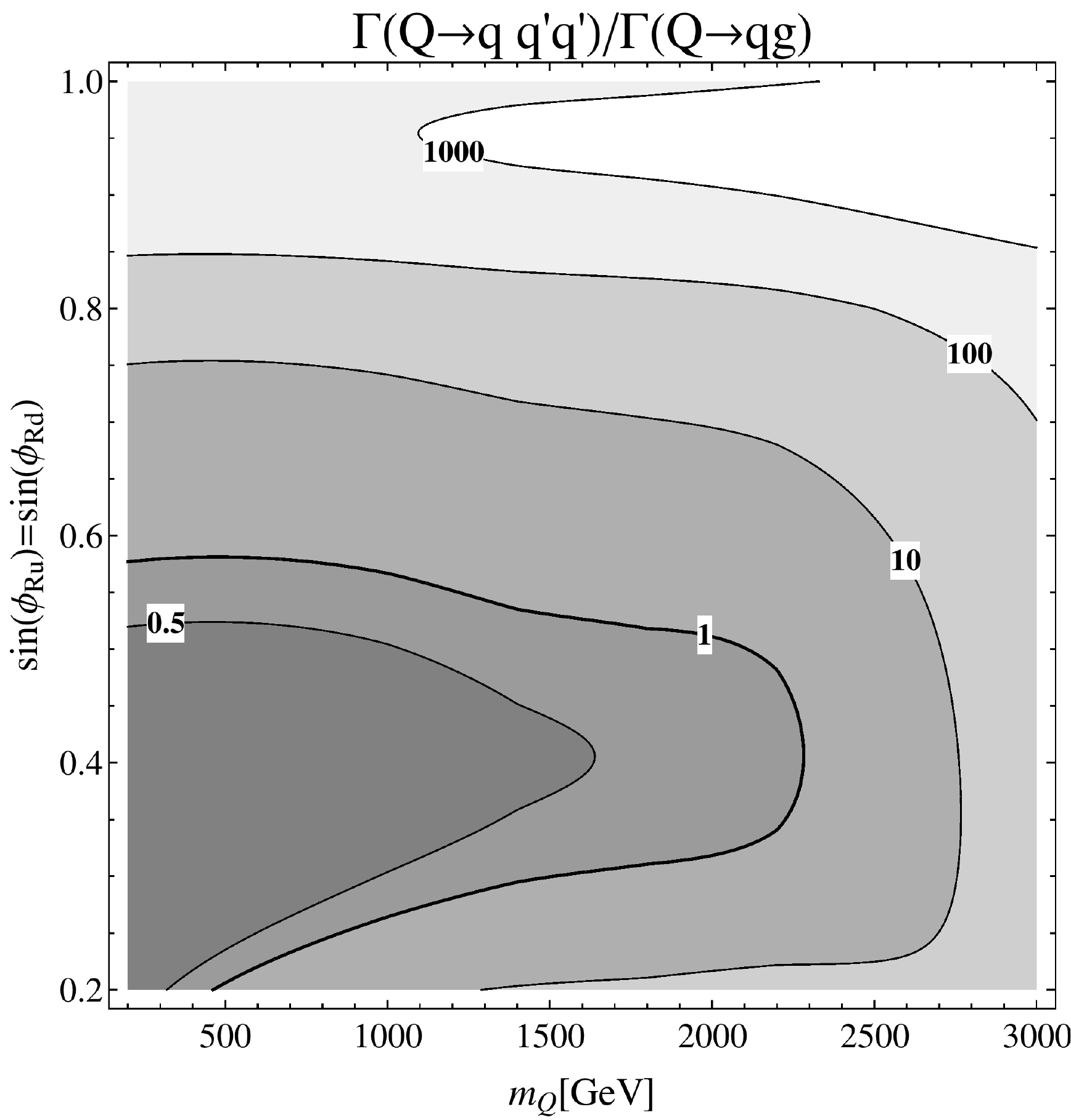}
		\caption{\emph{Relevance of the heavy fermion three body decay compared to the two body chromomagnetic decay displayed in the $(m_Q, \sin \phi_{Rq})$ plane for $m_\rho = 2.5$ TeV and $g_\rho = 3$. The thicker contour line indicates where the two and three body branching fractions are exactly equal to each other.}}
		\label{fig:fermionwidth}
	\end{center}
\end{figure}

The phenomenology and experimental strategies are strongly dependent on whether the two body or three body decay dominates, since this will result in either two or three jet final states. One interesting fact is that for $m_Q < m_\rho$ two body and three body decay scale in the same way with the masses. In figure \ref{fig:fermionwidth} it is shown in what regions of parameter space the two body or three body decay dominates. One should however keep in mind that other contributions could exist which possibly spoil this conclusion. Indeed the decay widths are in any case extremely small and so even normally subleading effects could be important.

\paragraph{Bounds on 3rd generation partners}:
Before delving into the direct searches of partners of the light generations let us consider the indirect bounds that can be derived from the top partners. Third generation partners behave very differently from light ones. Compositeness of left-handed top and bottom is sizable so even the right-handed partners decay through electroweak interactions as in the anarchic scenario. ATLAS places a strong bound on these states \cite{bib:atlas23}. The precise bound depends on the branching fraction (in the model under consideration the singlet $\tilde{T}$ decays in $W \, b$, $Z \, t$ and $h \, t$) but in any case it typically ranges between 500 and 700 GeV. If $\tilde{T}$ is at the bottom of the spectrum then the branching fractions can be predicted, $BR(\tilde{T}\to W b)\simeq 2 BR(\tilde{T}\to h t)\simeq 2 BR(\tilde{T}\to Z t)$, corresponding to an exclusion around 600 GeV.

In models that realize MFV this bound translates into a bound on the mass of light generation right-handed partners. This can only be avoided in extensions of MFV that allow to split the third generation \cite{bib:su2}. However this can only be done at the price of making the third generation partners heavier than the first two, at odds with naturalness. With this in mind we proceed to extract the direct exclusion limits that as we will see are rather weak.

\subsection{Single Production}
\label{sec:singleproduction}
If the heavy quark is singly produced in association with a light quark it then leads, at parton level, to three or four jet final states depending on whether chromomagnetic or color octet mediated decay prevails.

\subsubsection{Chromomagnetic Decay Scenario}
\label{sec:singleprodchromo}
The  topology of the event is  a pair of jets (originating from a quark and a gluon) with the invariant mass of the heavy partner and a third jet from the spectator quark. The bump hunter search of resonances decaying into dijets already considered in section \ref{sec:coloroctet} looks for features in the invariant mass of the two leading jets, where leading refers to $p_T$ ordering. Therefore this search will be effective if the heavy fermion is the father of the two leading jets, a situation that depends on $m_Q$ as we now explain. 

\begin{figure}[ht]
	\begin{center}
		\includegraphics[scale=0.42]{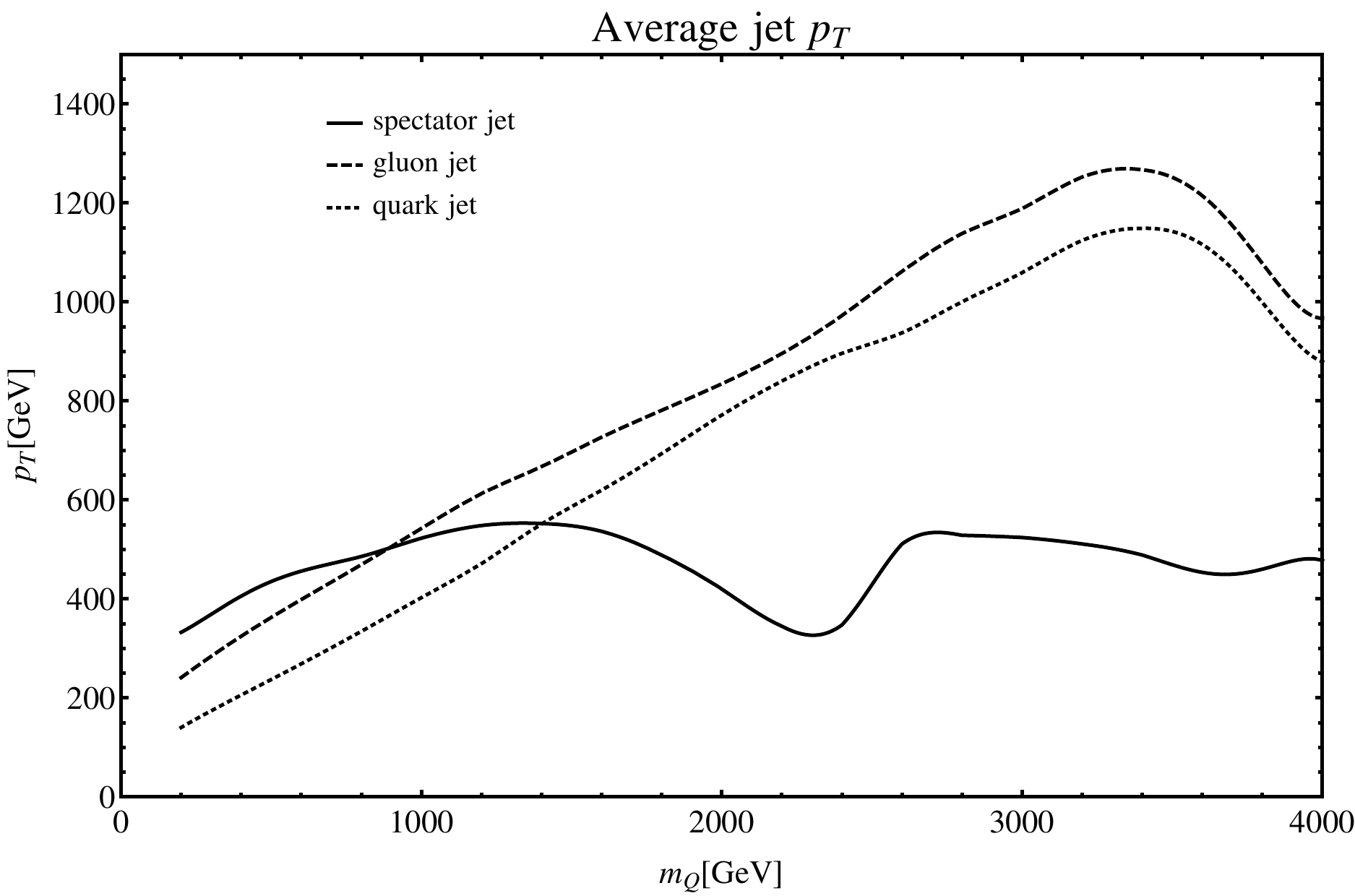} \hspace{0.2cm}
		\includegraphics[scale=0.42]{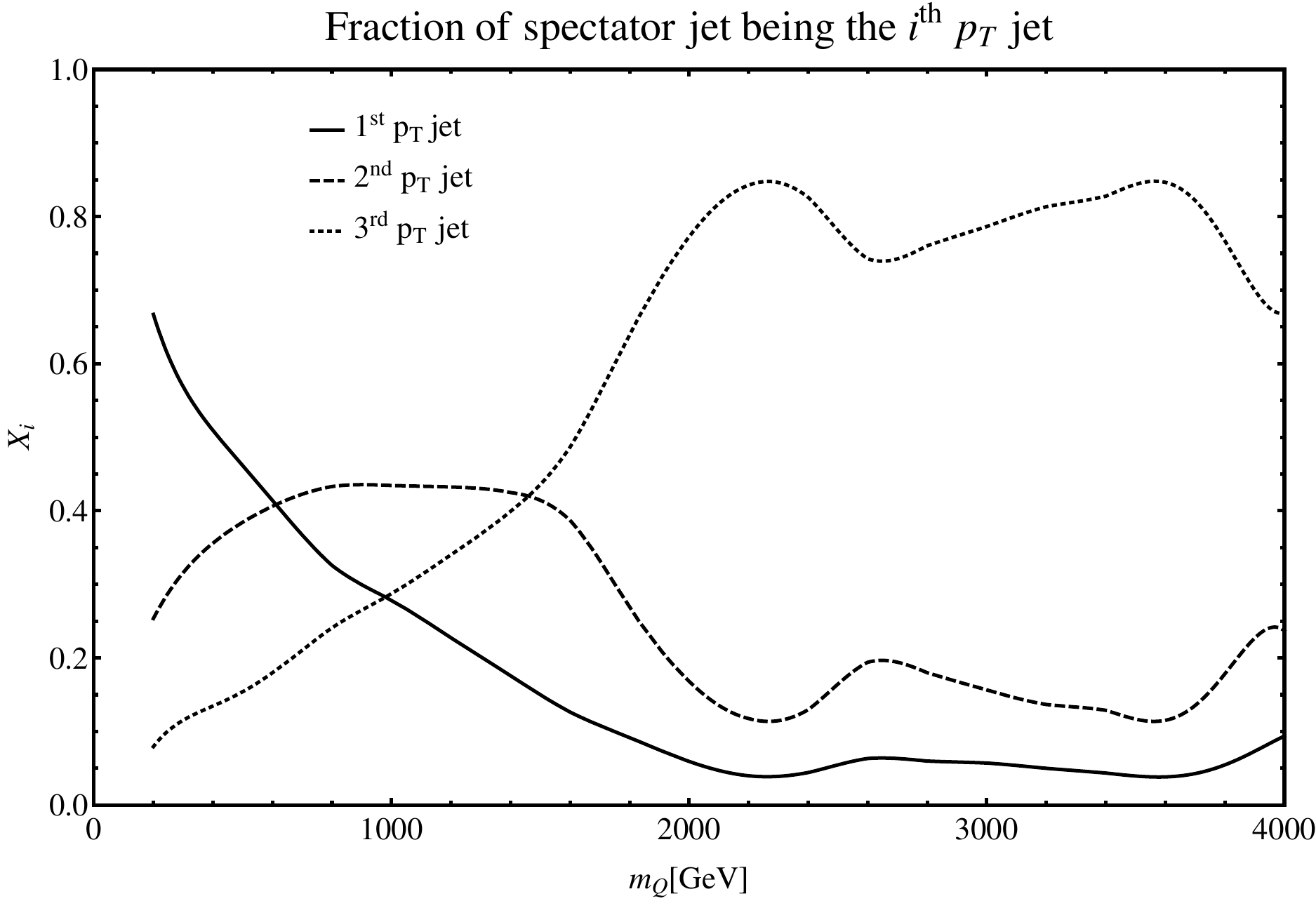}
		\caption{\emph{Plots of the $p_T$ spectrum of the three jet final state. On the left the average $p_T$ of the three different jets in the event as a function of $m_Q$ are displayed. On the right the fraction of events with the spectator jet being the $i^\mathrm{th}$ $p_T$ jet for LHC8.}}
		\label{fig:3jetptspectrum}
	\end{center}
\end{figure}

Let us first discuss the $p_T$ distribution of the recoiling jet due to the $t$-channel production of the heavy quark. One might think that the typical $p_T$ of the recoiling quark is controlled by the mass of the heavy fermion. In our region of parameter space however this  is not true due to parton distribution function (PDF) suppression. In fact we find that the average $p_T$ is almost independent on $m_Q$ being controlled by the total energy. This can be seen in figure \ref{fig:3jetptspectrum} on the left. The solid line is the average $p_T$ of recoiling quark obtained with $m_\rho=2$ TeV. For $E_{CM}=8$ TeV this is around 500 GeV. Moreover this feature persists for different values of $m_\rho$. In fact since as we have seen $m_\rho$ cannot be light, approximating the interaction with an effective operator is always a good approximation. Changing $m_\rho$ simply rescales the cross section. Quantitatively a good approximation to the cross section is given by 
\begin{eqnarray}
\frac{d\sigma}{d|p_T|} &\propto& \frac{1}{S}\, \frac{p_T^2}{
   m_\rho^4} \left(p_T+ \sqrt{m_Q^2+p_T^2}\right) \left( \frac{ p_T^2+m_Q^2+ p_T\sqrt{m_Q^2+p_T^2}}{S}\right)^{-\alpha} ,
\label{eq:dsigmadpT}
\end{eqnarray}
where $\alpha \sim 3-6 $ is a slowly varying function of $\hat s $ determined by the PDFs. This result is derived in appendix \ref{sec:ptordering}.

Given the $p_T$ of the recoiling quark we can derive the $p_T$ ordering of the jets in an event. Neglecting spin effects, the jets from the heavy quark will be  isotropically distributed in their CM frame with $p_T\sim m_Q/2$. Boosting to the lab frame one finds $p_T^{1,2}\sim \left|p_T^{spectator}\pm m_Q/2\right|$. Therefore for large $m_Q$ we expect the two jets to be leading and the opposite for small $m_Q$.

This is confirmed by our simulation. In figure \ref{fig:3jetptspectrum} on the right we plot the probability of the spectator quark to be the first, second or third jet in $p_T$. As we increase $m_Q$ the spectator quark tends to have the lowest $p_T$. Therefore in this region the standard dijet search will capture the signal. However, with the production cross sections given in figure \ref{fig:fermionprodcrossx} no bound is obtained in our model if we perform a recast. Moreover recent updates of dijet searches require a cut on the invariant mass of the jet pair to exceed $1$ TeV so this search is unlikely to produce a bound even in the future. The situation for $m_Q < 1.5 $ TeV is even less promising as in the case the spectator quark often gives rise to the first or second jet so the dijet search will not be efficient. In this case a different ordering of jets should be considered. Indeed requiring at least three jets in the final state and looking for bumps in the invariant mass of the second and third 
jet seems a promising strategy to reduce the background.

\subsubsection{Three Body Decay Scenario}
\label{sec:singleprodthreebody}
In this case we have a four jet final state with three jets reconstructing the mass of the heavy quark. The searches for four jets by ATLAS \cite{bib:atl4jets} and CMS \cite{bib:cms4jets} with no missing energy cuts are optimized for pair production of a heavy resonance, both decaying into two jets. Although these searches share the same final state, they have a low efficiency to pick up our signal because of the different topology.
\begin{figure}[ht]
	\begin{center}
		\includegraphics[scale=0.42]{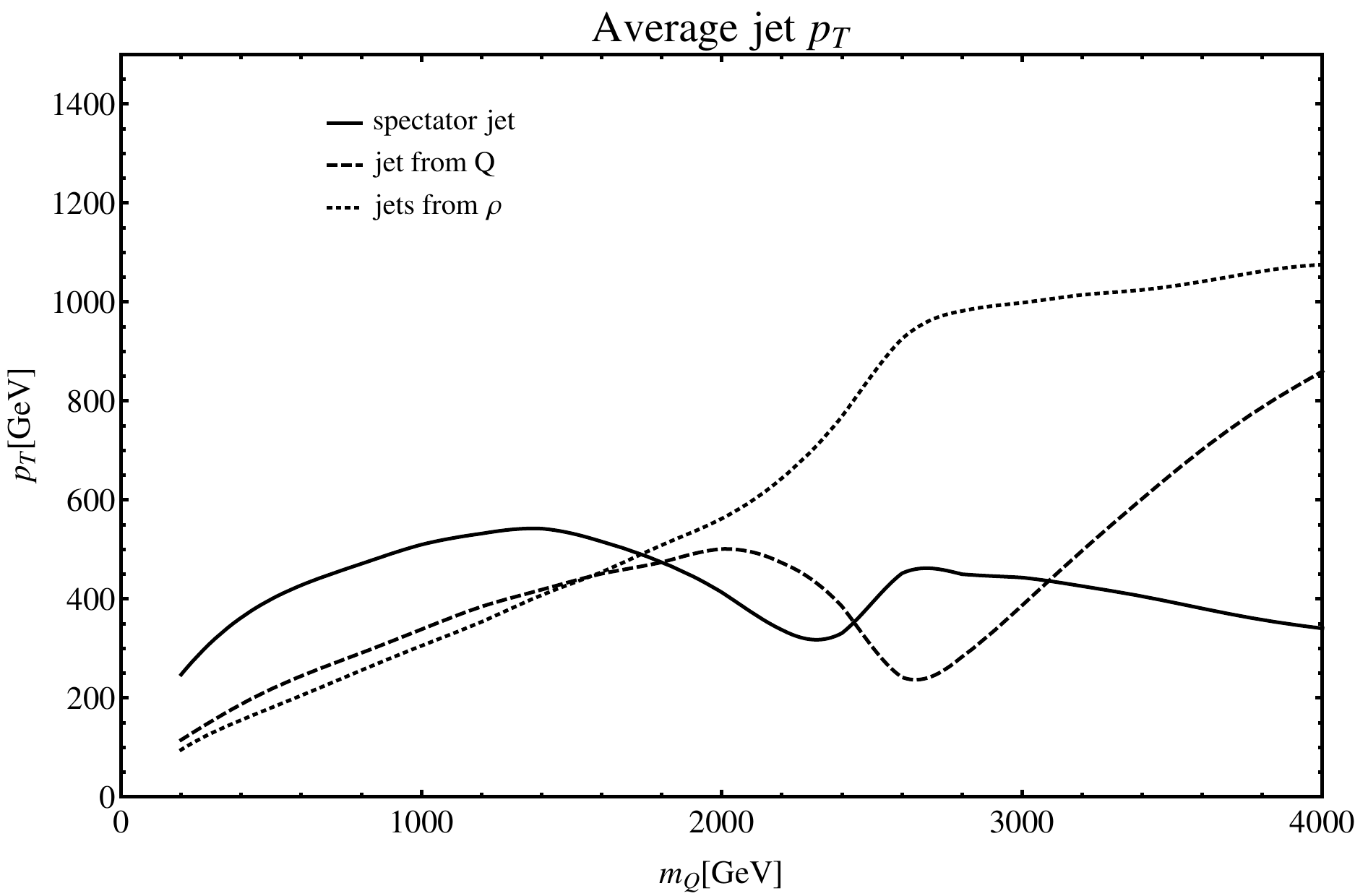} \hspace{0.2cm}
		\includegraphics[scale=0.42]{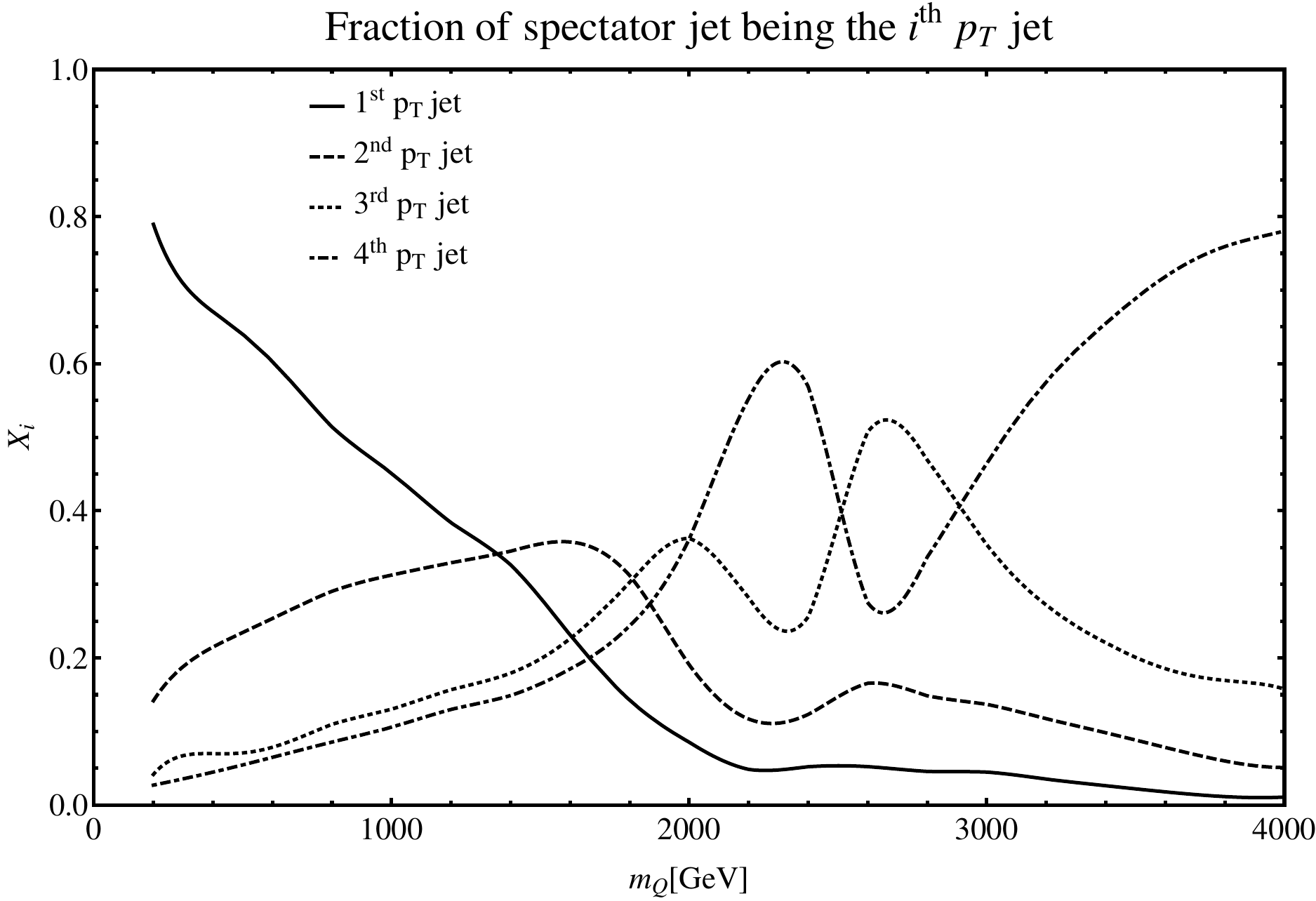}
		\caption{\emph{Plots of the $p_T$ spectrum of the four jet final state. On the left the average $p_T$ of the four different jets in the event as a function of $m_Q$ are displayed. On the right the fraction of events with the spectator jet being the $i^\mathrm{th}$ $p_T$ jet for LHC8.}}
		\label{fig:4jetptspectrum}
	\end{center}
\end{figure}

Obviously to capture the signal one should study the invariant mass of trijets. As previous case an important element is the ordering of jets. This is shown in figure \ref{fig:4jetptspectrum} on the right. Qualitatively this is similar to the two body decay. As intuitive however the recoiling jet is more likely to be the leading jet. This happens 50\% of the times for a fermion with 1 TeV mass. Therefore in this case a dedicated search pairing the second, third and fourth jet is expected to be very effective.

\subsection{Double Production}
\label{sec:doubleproduction}
We have in this case 4 or 6 jet final states at parton level. 5 jets could also be obtained in certain regions of parameters where 2 body and 3 body decay are comparable but we will neglect this possibility. In the 4 jets case two pairs of jets form the same invariant mass equal to the heavy quark mass. In the 6 jets case two sets of three jets each form the invariant mass of the heavy quark. 

\subsubsection{Chromomagnetic Decay Scenario}
\label{sec:doubleprodchromo}
CMS and ATLAS analyzed double dijets final states, where they look for pair production of a heavy resonance decaying into two jets in \cite{bib:atl4jets,bib:cms4jets}. ATLAS only considers a mass region  between $150$ and $350$ GeV, whereas CMS considers a region from $320$ to $1200$ GeV. Since our interest is mainly in the mass region up to around $1$ TeV for the heavy quark partners, only the CMS analysis is considered. This search is expected to be effective for relatively low partner masses since for high masses the three body decay is favoured, see figure \ref{fig:fermionwidth}.

The CMS analysis investigates events with at least four jets with $|\eta| < 2.5$ and $p_T > 150$ GeV and then combines the four highest-$p_T$ jets into dijet combinations with $\Delta R_{jj} > 0.7$. Then the dijet pair combination with minimal $\Delta m / m_\mathrm{avg}$ is selected, where $\Delta m = | m_\mathrm{jj}^{(1)} - m_\mathrm{jj}^{(2)} |$ and $m_\mathrm{avg} = \tfrac{1}{2} ( m_\mathrm{jj}^{(1)} + m_\mathrm{jj}^{(2)} )$, with a maximum $\Delta m / m_\mathrm{avg} < 0.15$ to suppresses the QCD background. Then a last requirement is
\begin{equation}
	\Delta = \sum_{i=1,2} p_{T,i} - m_\mathrm{avg} > 25 \, \mathrm{GeV} ,
\end{equation}
ensuring a smoothly falling paired dijet mass spectrum. In the absence of any observed resonances CMS then provides the limits on the folded $\sigma \times \mathrm{Br} \times \epsilon$ as a function of the resonance mass, to which our scenario will be compared to obtain limits.

\begin{figure}[ht]
	\begin{center}
		\includegraphics[scale=0.46]{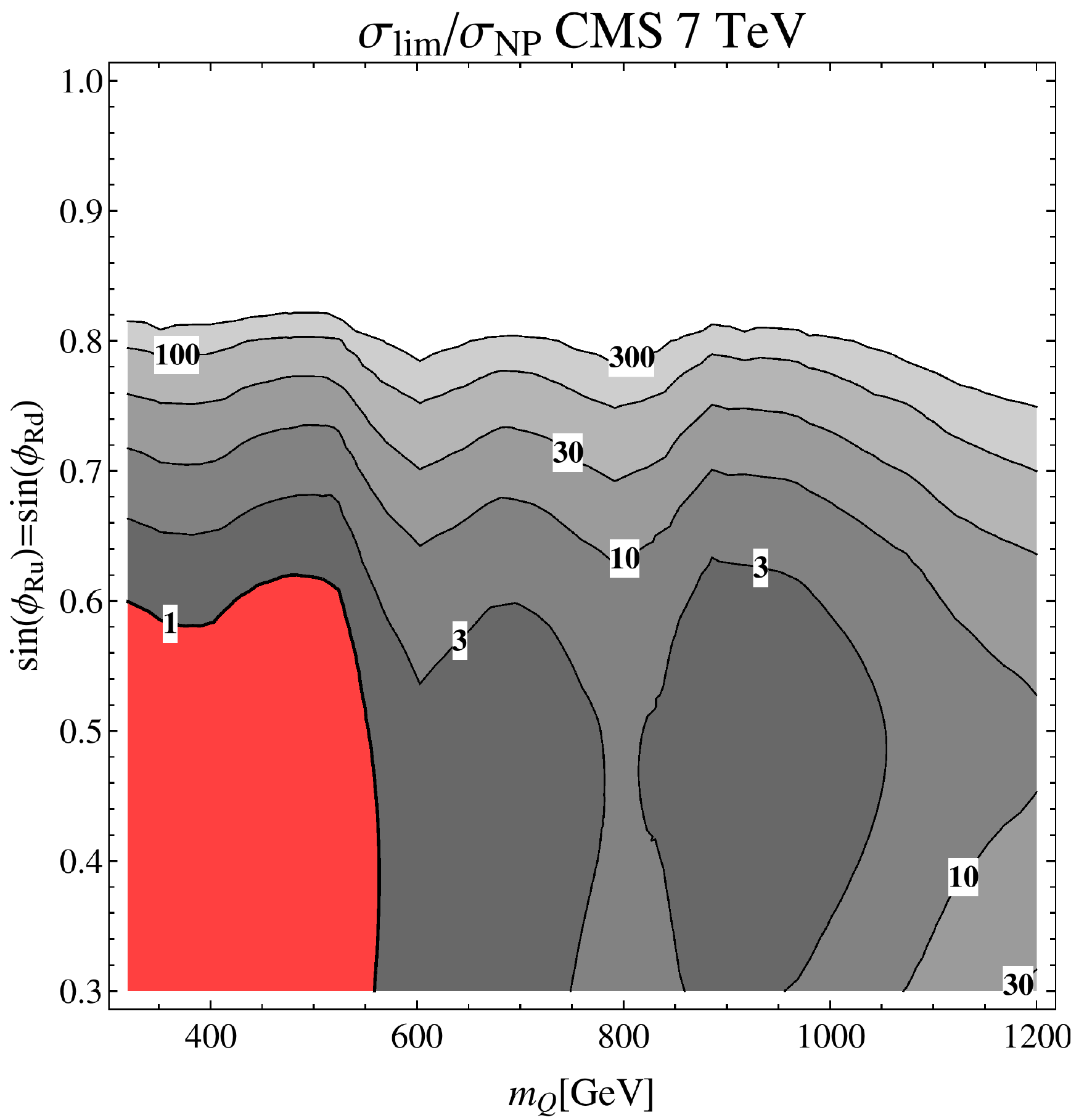}
		\caption{\emph{Constraints from the CMS double dijet search \cite{bib:cms4jets} for the double production combined with chromomagnetic decay scenario for $g_\rho = 3$ and $m_\rho = 2.5$ TeV. Displayed is the limit cross section over the new physics cross section in the $(m_Q, \sin \phi_R)$ plane. The red region is already excluded at $95$\% CL by the current searches, whereas the gray contours give an indication for the needed increase in sensitivity to exclude further regions. The peculiar shape, the ``island'' in particular, is accounted for by upward fluctuations in the data around $600$ and $800$ GeV.}}
		\label{fig:doubledijet}
	\end{center}
\end{figure}

Our scenario is almost completely equivalent to the coloron model considered in the CMS analysis only differing in the production modes. Therefore we expect similar final state topology and the selection criteria to be next to optimal. To compare with the coloron exclusion limits we generate the dijet resonances using our FeynRules-MadGraph-Pythia-Delphes chain  (also in our case the width of the resonance is negligible compared to the experimental resolution). For a set of points in the $(m_Q, \sin \phi_R)$ plane we analyze the efficiencies and obtain a value for $\sigma \times \mathrm{Br} \times \epsilon$ to be compared to the CMS limit. We focus on the excluded region  and the possible exclusion potential. Therefore we plot the limiting cross section $\sigma_\mathrm{lim}$ divided by the new physics cross section $\sigma_\mathrm{NP}$ of our model which removes the dependence on branching ratios and acceptances. This gives a good indication of the increase in sensitivity required to exclude certain regions 
of parameter space. The resulting contour plot is given in figure \ref{fig:doubledijet}. Any region with $\sigma_\mathrm{lim} / \sigma_\mathrm{NP} \leq 1$ is excluded by the current searches, this is the red contour with the thick edge. We conclude that heavy partners with masses between $320$ and $500$ GeV are excluded, provided that the chromomagnetic decay dominates. 

\subsubsection{Three Body Decay Scenario}
\label{sec:doubleprodthreebody}
\begin{figure}[ht]
	\begin{center}
		\includegraphics[scale=0.46]{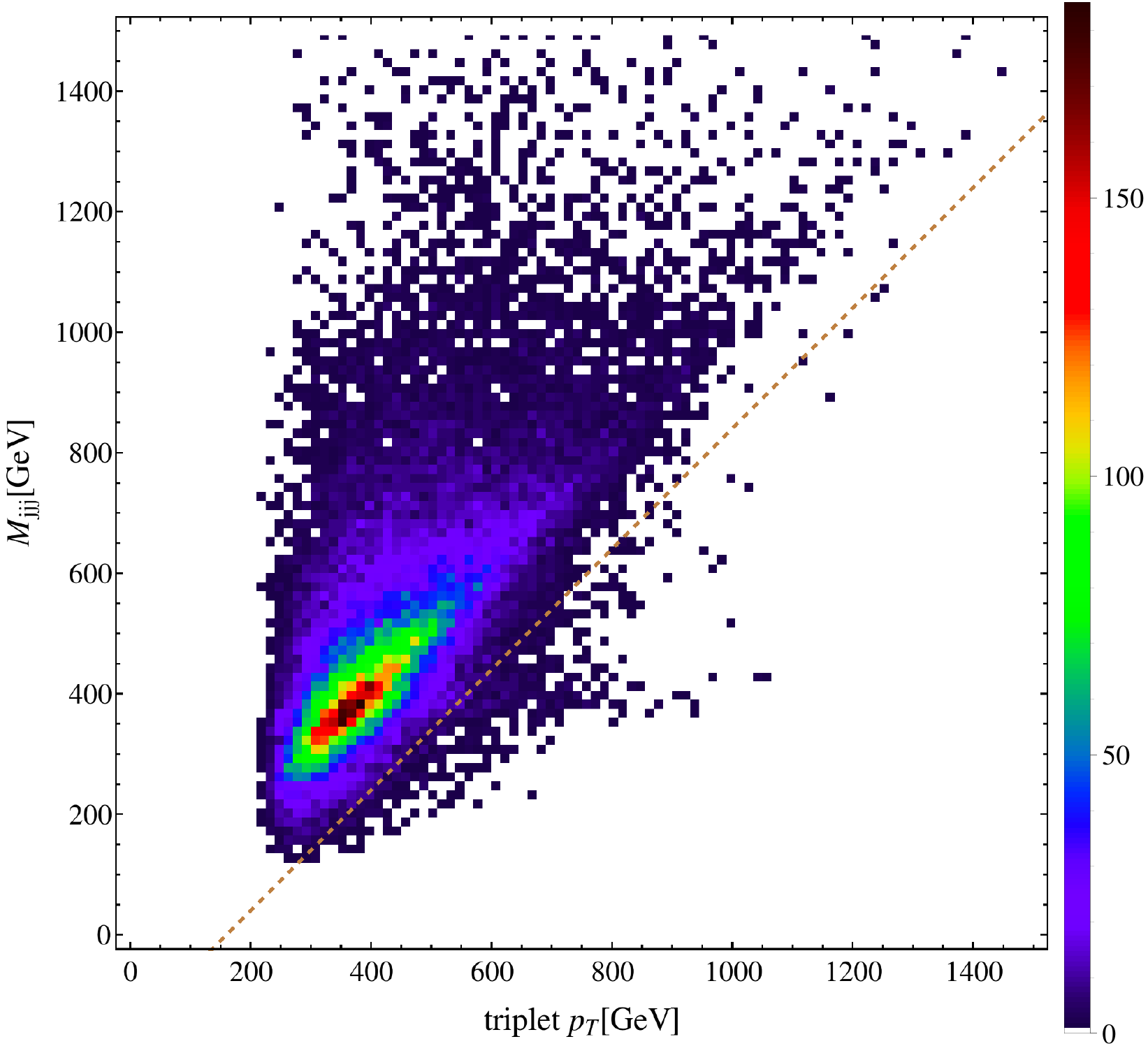} \hspace{.2cm}
		\includegraphics[scale=0.46]{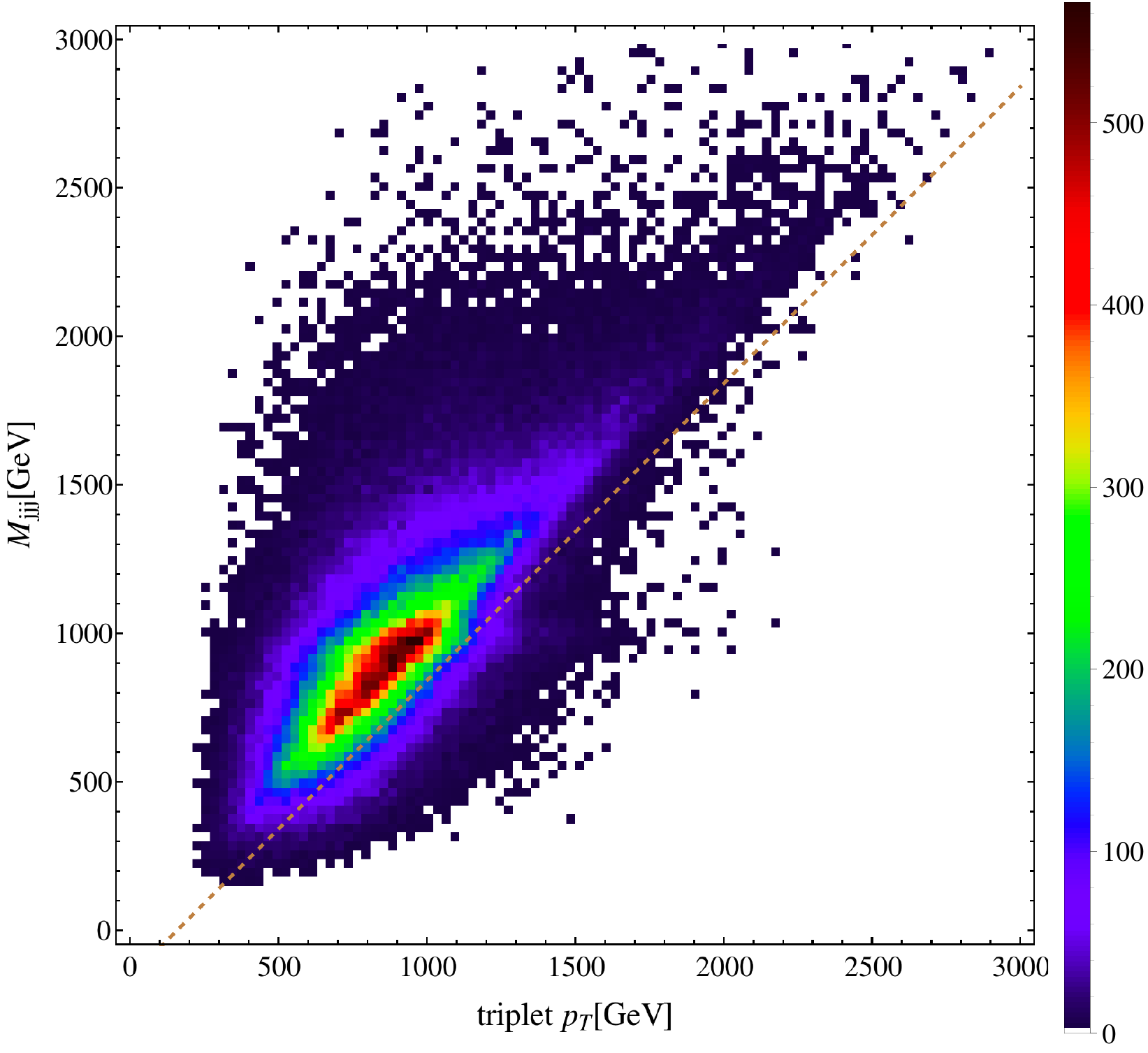}
		\caption{\emph{Distribution of triplets mass $M_{jjj}$ versus the triplet scalar $p_T$ of all 20 triplets in each event for LHC7. 	For quark partner masses of $400$ GeV (left) and $1000$ GeV (right), the selection criterion from equation \eqref{eq:cms6jselection} is given by the orange dashed line. These plots give an indication of the leakage of combinatorial background into the signal region.}}
		\label{fig:tripletdistribution}
	\end{center}
\end{figure} \noindent
In this case we have a six jet final state, where two combinations of three jets originate from identical mother particles. The search closest to this topology we are aware of is by CMS \cite{bib:cms6jets} where they look for the invariant mass of three jets in events with at least six jets with $p_T > 70$ GeV and $|\eta| < 3.0$. Furthermore the total scalar sum of $p_T$ is required to be higher than $900$ GeV for each event. The search aims to capture  pair produced trijet resonances and is interpreted in terms of RPV gluinos decaying into three jets. The six highest $p_T$ jets are combined into all 20 three jet combinations and in order to reduce both combinatorial and QCD background the requirement
\begin{equation} \label{eq:cms6jselection}
	M_{jjj} < \sum_{i=1}^3 p_T^i - \Delta \, , \quad \left( \Delta = 160 \; \textrm{GeV}\right)
\end{equation}
for each triplet is imposed. The acceptance is then defined as all the events with at least one triplet of jets passing this cut. The experiment provides the $95$\% CL limits on $\sigma \times \textrm{Br}$ as a function of the resonance mass in the range from $280$ GeV up to $1000$ GeV.

Even though our topology shares the same final state as the RPV gluinos studied in the CMS analysis the kinematics are quite different. One of the main differences is that the quark partners are produced mostly by the color octet rather than the gluon. Moreover the gluino decay is modeled by a four fermion effective interaction, whereas the heavy quark decay proceeds through an off-shell color octet. Hence, for the analysis to be applicable roughly the same acceptances for both scenarios should be obtained. Especially the selection criterion in equation \eqref{eq:cms6jselection} should have the same effect on the combinatorial background and the trijets coming from the decay of the heavy partner. This is relevant since after this selection criterion a resonance search in the triplet invariant mass spectrum is performed. In order to analyze the effect of the cuts, in particular \eqref{eq:cms6jselection}, our signal has been simulated and the distribution of events in the trijet mass versus triplet scalar $p_
T$ has been plotted in figure \ref{fig:tripletdistribution}.

\begin{figure}[ht]
	\begin{center}
		\includegraphics[scale=0.5]{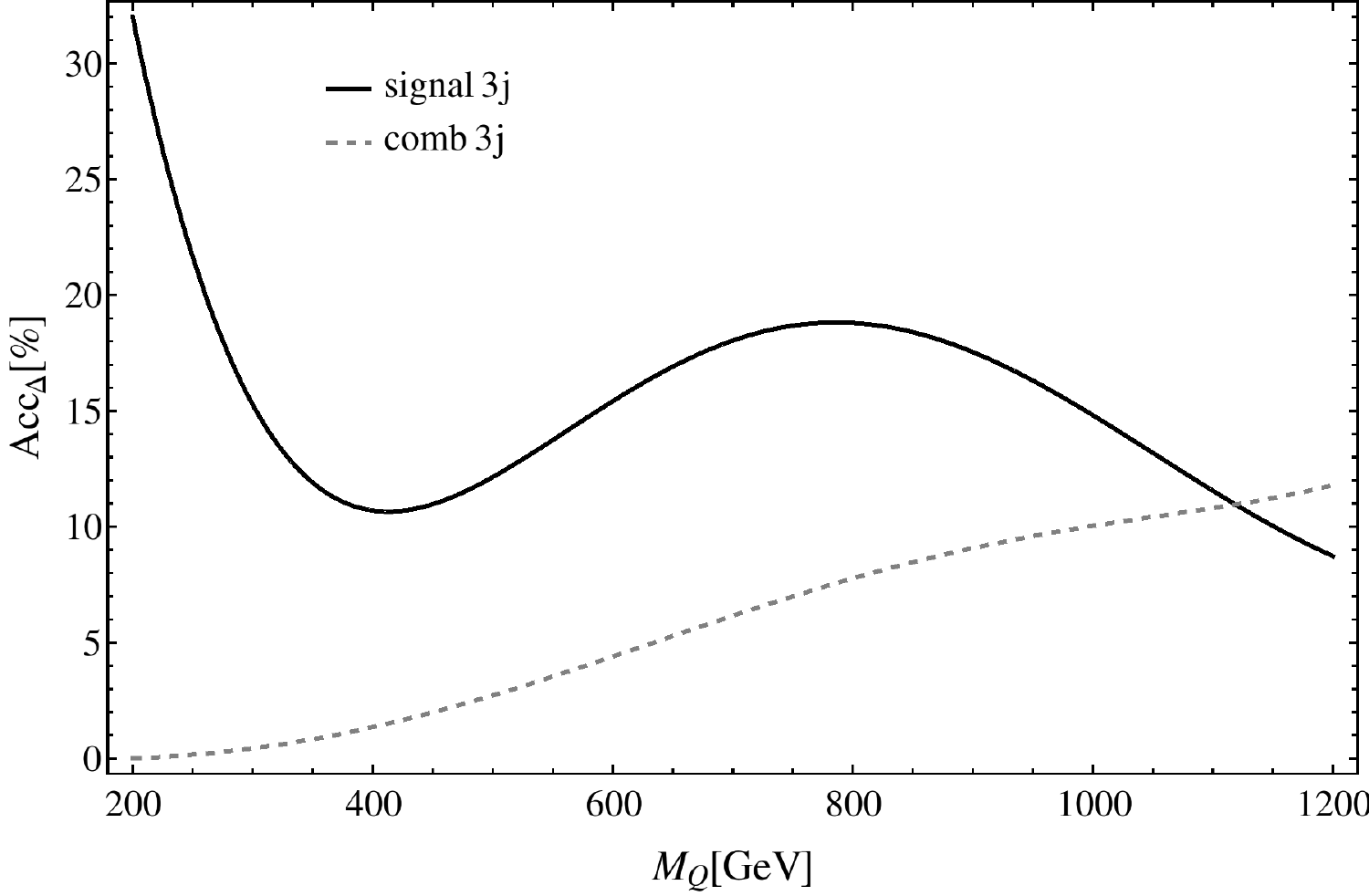}
		\caption{\emph{Acceptance of triplet events for the selection criterion in equation \eqref{eq:cms6jselection}. The black line shows the acceptance for the triplets originating from one of the heavy quarks, whereas the gray dashed line shows the acceptance for the other triplets forming combinatorial background.}}
		\label{fig:tripletdeltacut}
	\end{center}
\end{figure}

The CMS analysis is optimized for the gluino scenario choosing $\Delta = 160$ GeV and for triplets originating from the gluino the probability for passing this selection criterion ranges between $2$\% and $13$\% depending on the gluino mass. In figure \ref{fig:tripletdeltacut} the acceptance for our signal is plotted, from which one can see that the acceptances are generally higher, however also the combinatorial background grows. At high masses we see that the combinatorial background starts to dominate over the signal, hence the selection as in equation \eqref{eq:cms6jselection} is not efficient for our topology. Therefore no significant bound can be extracted. However, the search for this final state is potentially interesting and could be optimized with minor effort for the topology of right-handed compositeness.

\section{Dedicated Searches}
\label{sec:dedicatedsearches}
In the previous section we recasted the existing multi-jet searches of ATLAS and CMS to set limits on the heavy quark partners. Those limits are rather weak for the right-handed partners, see for example figure \ref{fig:doubledijet}, since the searches are not optimized to the most distinctive topology of the model, the single production of the heavy quark $Q$ in association with a light jet. In this section we are going to propose searches that exploit the characteristic behavior of this production mechanism, namely $p p \to Q q$. This leads to 2+1 (3+1) jets for the two body (three body) decays of the heavy quark partners.

\begin{figure}[ht]
	\begin{center}
		\includegraphics[scale=0.24]{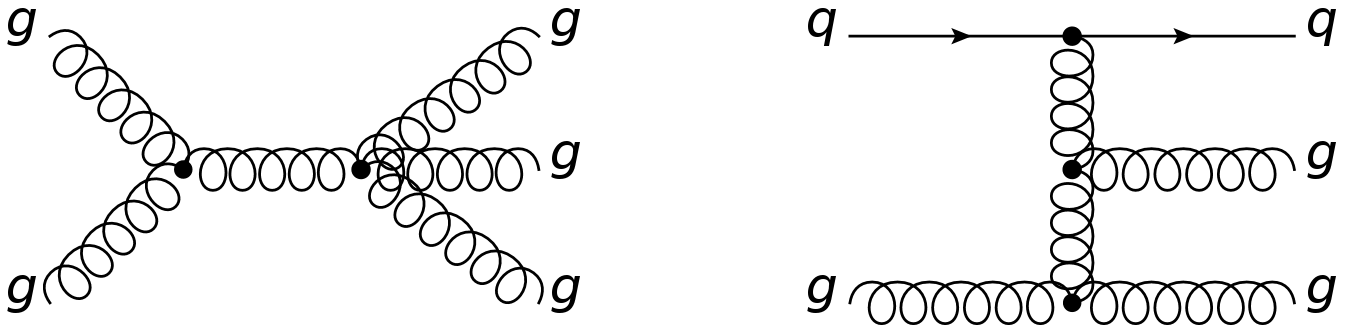}
		\caption{\emph{Typical QCD background events leading to three high-$p_T$ jets.}}
		\label{fig:qcd3j}
	\end{center}
\end{figure}

The topology is characterised by at least three hard jets, where some of the jets reconstruct the mass of the fermion $Q$. 
The main background arise from QCD jets and is dominated by the diagrams given in figure \ref{fig:qcd3j}.

When looking at the dijet searches, a cut on the hardness of the third and fourth jet aids in reducing the background, but the effect is not drastic. For example, in 8 TeV QCD samples where the two leading jets have $p_T>$ 150 GeV, asking for a third one with $p_T >$ ($25$, $70$, $100$ and $150$) GeV has an efficiency of ($40$, $9$, $4$ and $1$)\%. A larger reduction of the background can be achieved with more sophisticated cuts, for which we provide details later.

Other studies with some overlap with the single production topology are the CMS and ATLAS studies of double dijets. In the previous section, we applied one of these searches to double production $p p \to Q Q$ where $Q\to j j$, concluding that the reach is rather weak. In the single production case the sensitivity is even lower, as the topology does not resemble the double dijet. For example, the efficiencies of the signal $p p \to q Q$ where $Q\to 3 j$ to this search for $m_Q=$ $1$, $2$ and $3$ TeV is in the range of $1$-$4$\%. As in the case of dijet bump searches, a dedicated search should be carried out. In the following we discuss the kinematic variables which show a better discrimination power of signal versus background for this topology. Two benchmark scenarios are considered which correspond to $m_\rho = 2500$ GeV, $g_\rho = 3$ and $\sin \phi_{R_{u,d}} = 0.6$, where the heavy fermion partner mass equals $m_Q = 600$ or $1200$ GeV. The relevant kinematic variables are now discussed.

\begin{figure}[ht]
	\begin{center}
		\includegraphics[scale=0.36]{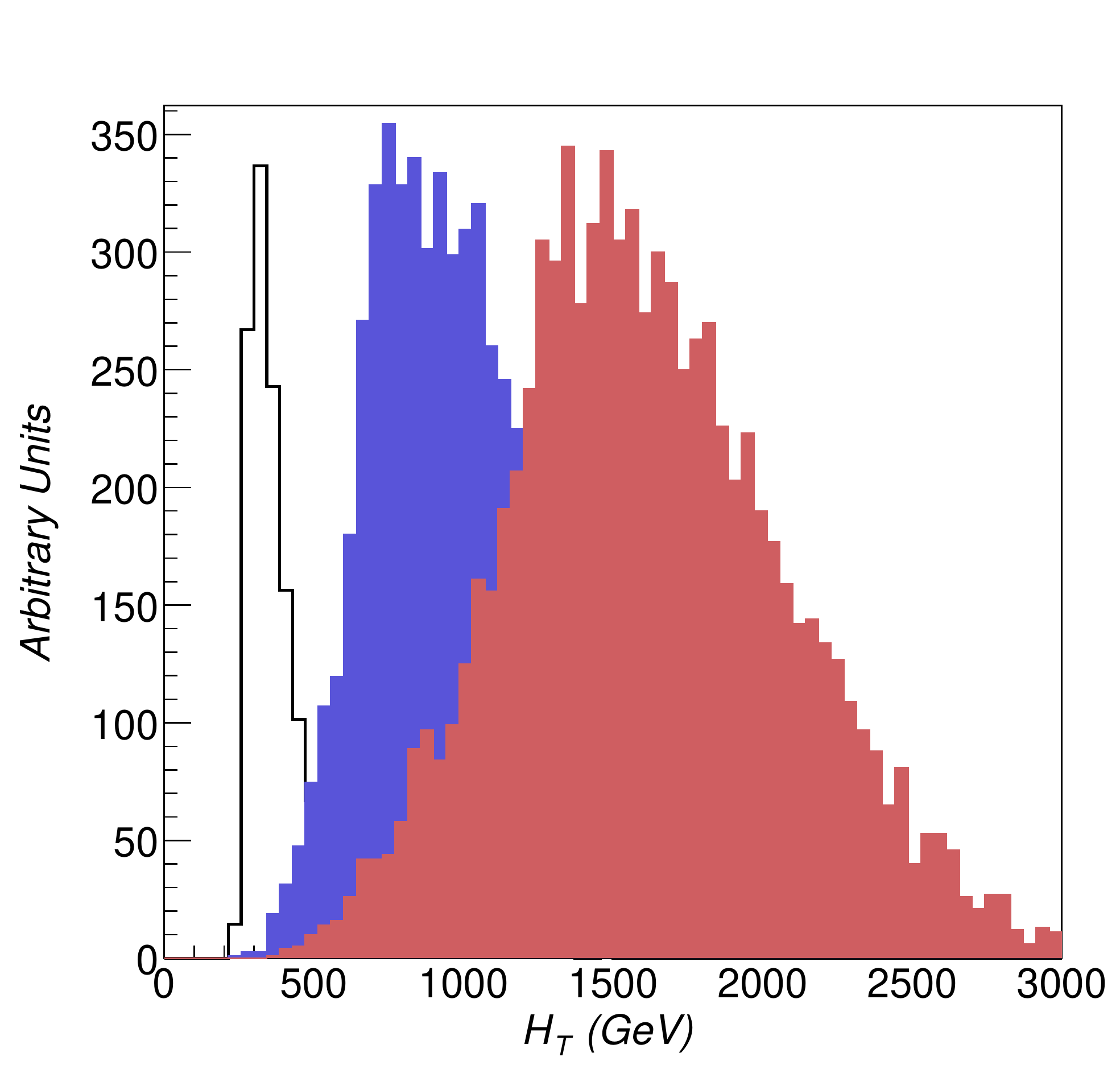}
		\includegraphics[scale=0.36]{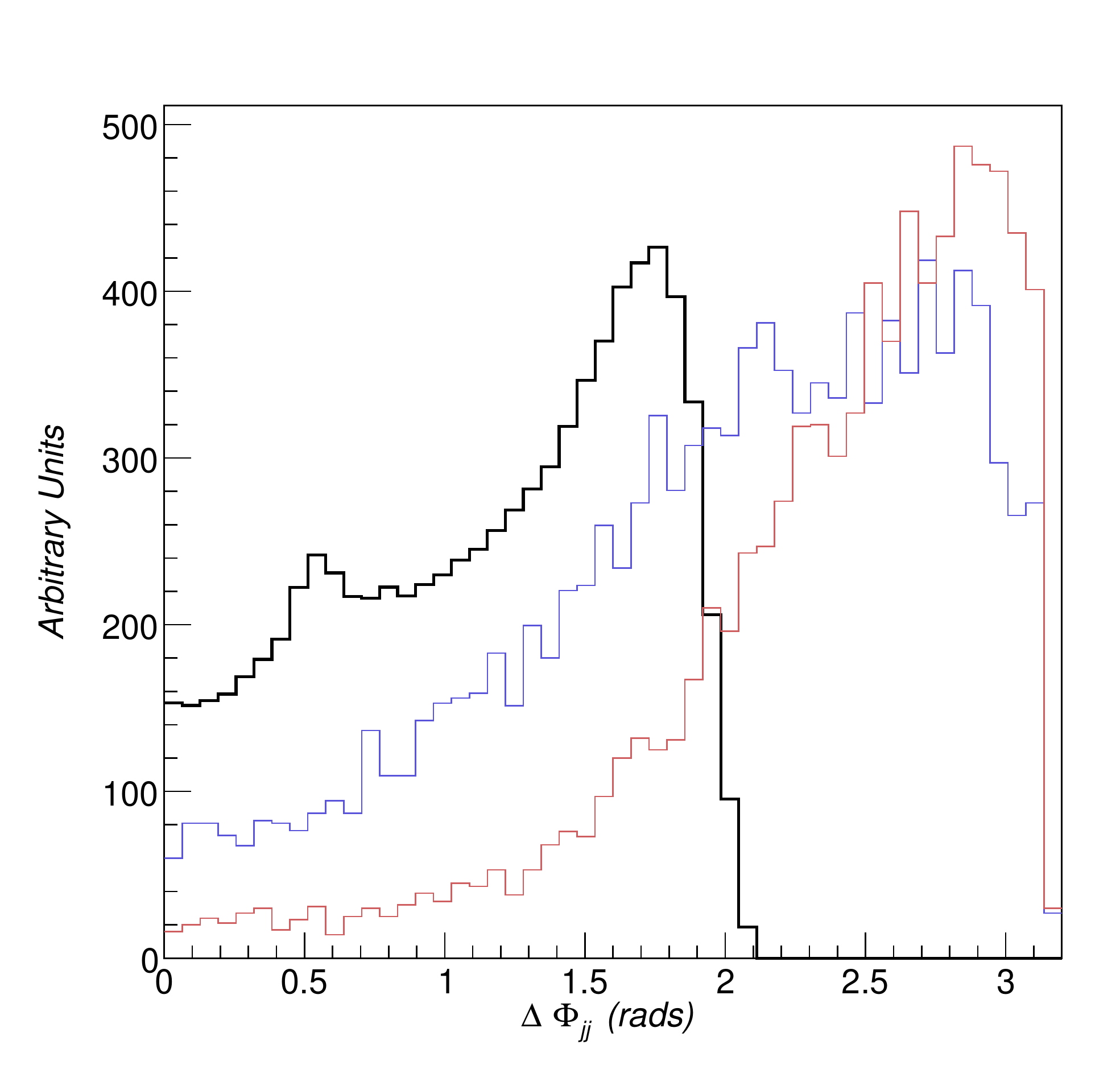}
		\caption{\emph{{\bf (Left) } The $H_T$ distribution for QCD events with $n_j \geqslant 3$ (black line), 2+1 signals with $m_Q=$ 600 GeV (purple distribution) and $m_Q=$ 1200 GeV (magenta distribution) in LHC8. {\bf (Right)} The $\Delta \phi$ distributions between the two subleading QCD jets (black) and the dijets from the decaying $Q$ particle, with $m_Q=$ 600 (1200) GeV in blue (red). Both figures are generated at parton level, and in the right figure truth information is used to identify the jets from the heavy quark. }}
		\label{fig:Htcomp}
	\end{center}
\end{figure}

\paragraph{The $H_T$ variable:}
We use the usual definition of the $H_T$ variable
\begin{equation}
H_T=\sum_{i=\textrm{jets}} p_{T,i} \, .
\end{equation}
In figure \ref{fig:Htcomp} (left) one can see that signal has a larger $H_T$ distribution than the background, and it increases with $m_Q$. Note that in this plot and all the following ones, basic cuts on the jets are $p_{T,j}>$ 70 GeV and $|\eta_j|<$ 2.5.

\paragraph{Angular distribution:}
Since the heavy resonance is produced with little boost, one would expect a symmetric angular distribution among the jets coming from the decaying particle. In the  $Q\to 2j$ case, the daughter jets tend to be produced with $\Delta \phi = \pi$, whereas in the $Q\to 3j $ case one would expect a distribution near $\Delta \phi =2 \pi/3$.

\begin{figure}[ht]
	\begin{center}
		\includegraphics[scale=.46]{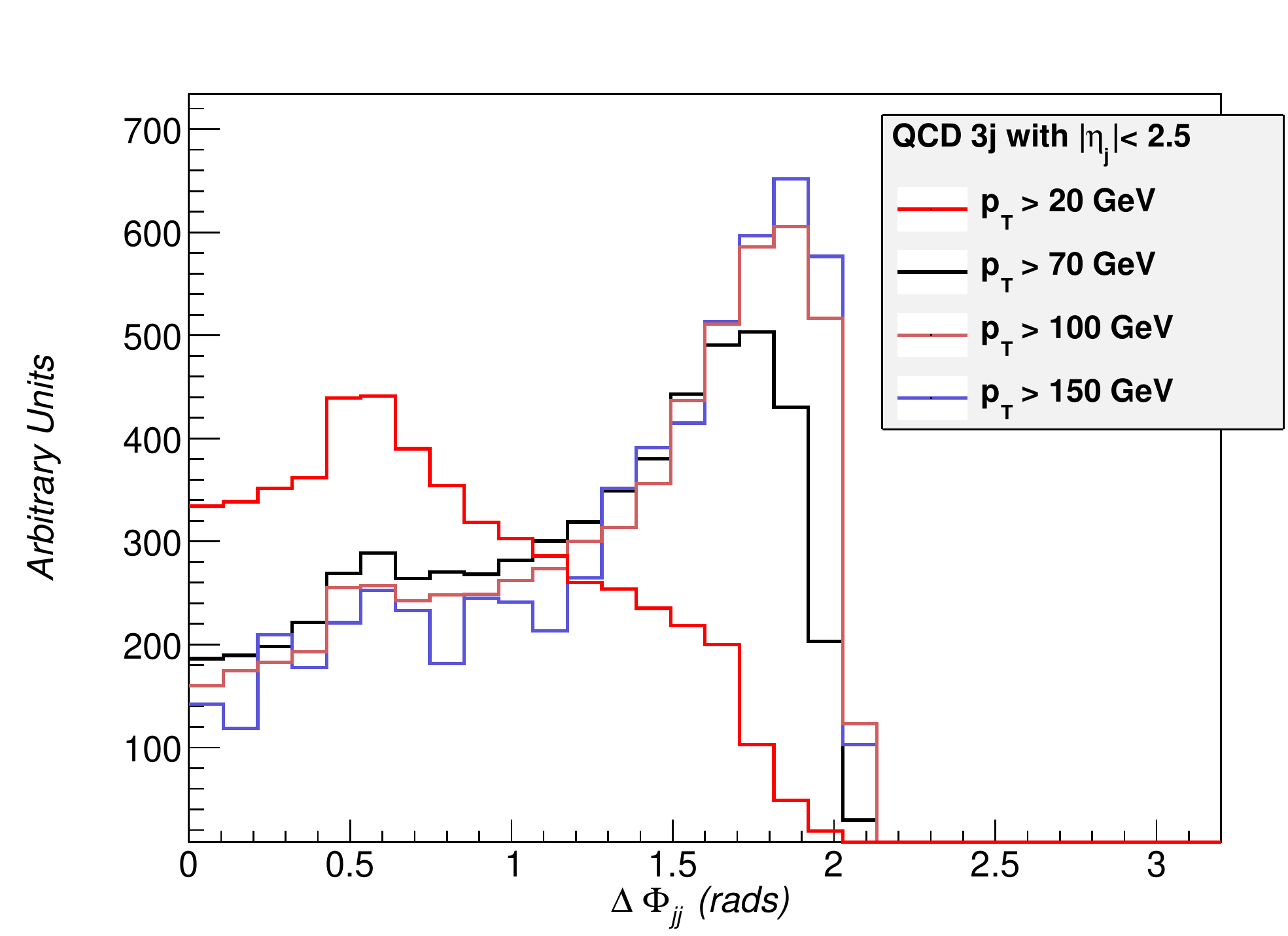}
		\includegraphics[scale=.36]{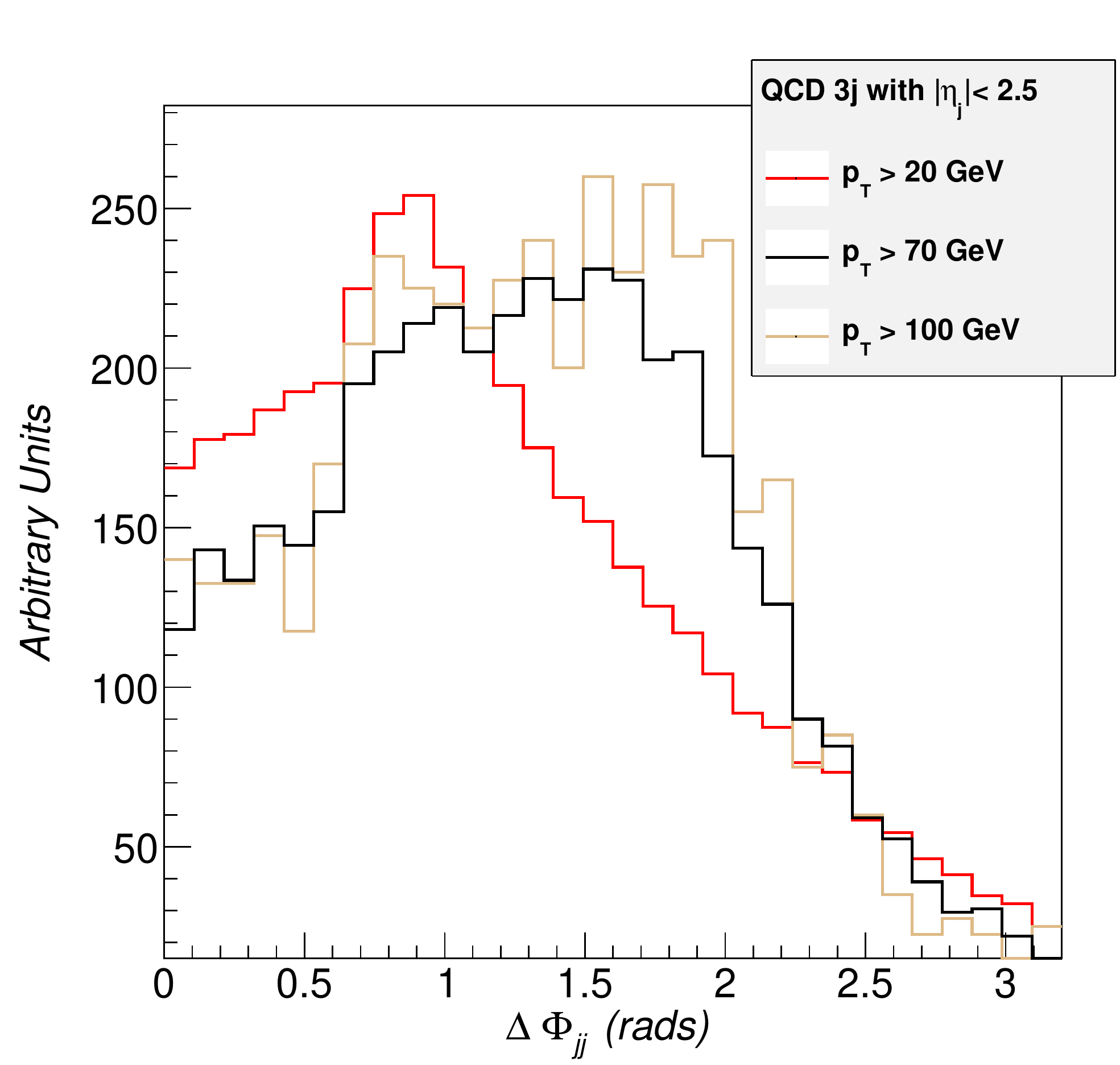}
		\caption{\emph{The angular distribution between the two subleading jets in QCD events, as a function of the $p_{T}$ cuts on all jets for LHC8. The left plot is the partonic result and the plot on the right is reco-level.}}
		\label{fig:QCDpT}
	\end{center}
\end{figure}

In the QCD case, though, jets would not have such a preference. In three jet QCD events, like the ones in figure \ref{fig:qcd3j}, one would expect a rather symmetric distribution of jets, more so as we increase the cut on $p_T$. This is seen in figure \ref{fig:QCDpT}, where as we increase the $p_T$ threshold, the distribution is more and more peaked towards $2 \pi/3$, hence the {\it Mercedes} configurations. With the same cut on all jets, this configuration minimizes the overall centre of mass energy of the three jet system, $M \sim 3 p_{T,min}$. For the configuration where the subleading jets are close and back-to-back with the leading jet, the minimal mass equals $M \sim 4 p_{T,min}$. Here the two subleading jets have $p_T=p_{T,min}$ and the leading jet $p_T$ is $2 p_{T,min}$ to balance momentum.

This is shown in figure \ref{fig:Htcomp} (right), where we see that the jets from the $Q$ decay tend to be symmetric, more so as the mass increases and the $Q$ has smaller boosts. We find similar discriminating features when looking at the $Q\to 3 j$ case, now with the peak at $\Delta \phi = 2 \pi/3$ for the three jets from $Q$. Note, though, that figure \ref{fig:Htcomp} (right) has been done using parton level truth events (where the information of the mother particle was known). When showering, detector effects and combinatorial background is added, the discriminating power of $\Delta \phi_{jj}$ is greatly reduced.
 \begin{figure}[ht]
	\begin{center}
		\includegraphics[scale=0.36]{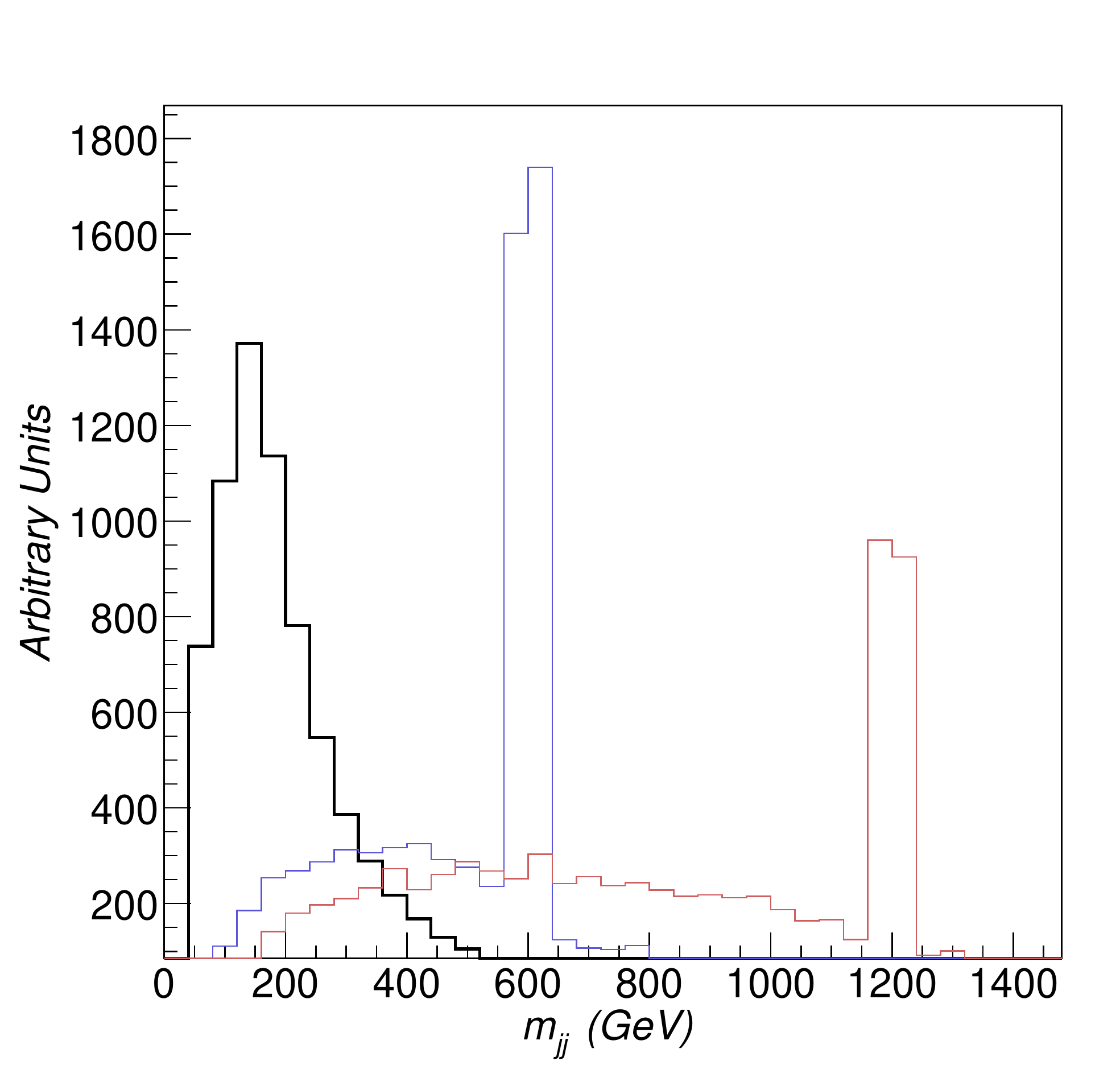}
		\includegraphics[scale=0.36]{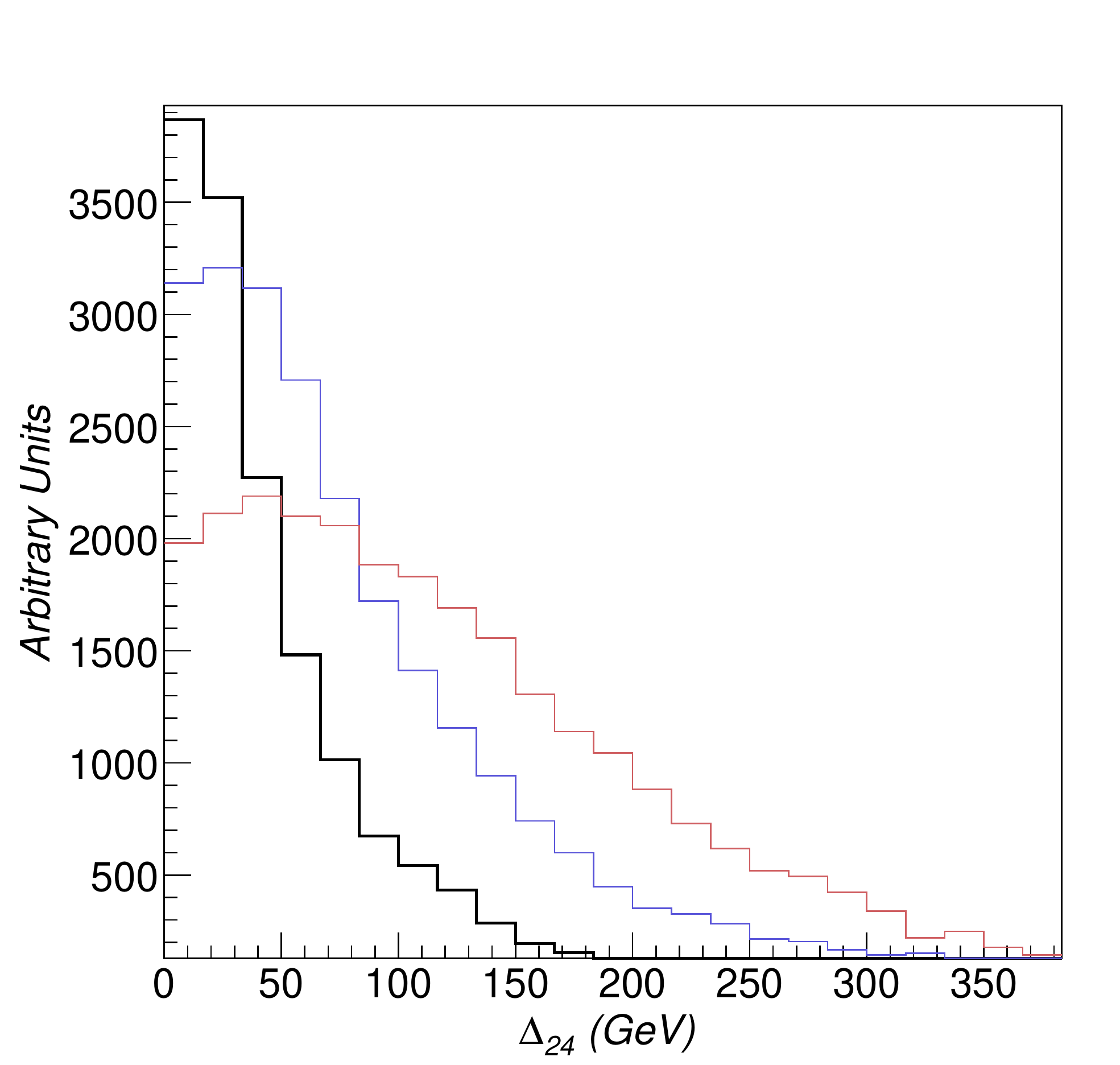}
		\caption{\emph{ {\bf (Left) } The invariant mass distribution of the two subleading jets for QCD events (black) and signal events in the 2+1 topology with $m_Q=$ 600 (1200) GeV in blue (red) at LHC8.  {\bf (Right) } The distribution $\Delta_{24}$ for QCD events (black) and signal events in the 3+1 topology with $m_Q=$ 500 (1000) GeV in blue (red) at LHC8. Both plots are generated at the parton level.}}
		\label{fig:mjjsub}
	\end{center}
\end{figure}

\paragraph{Mass bump reconstruction:}
An obvious characteristic of the signal is the presence of a mass bump, {\it if} the right combination of jets was chosen. In the previous section, we showed in figures \ref{fig:3jetptspectrum} and \ref{fig:4jetptspectrum} (right panels), that the leading jet tends to be the spectator jet for low $m_Q\lesssim$ 1 TeV, more so for the 3+1 than the 2+1 topology. We then choose in each event the two (three) subleading jets and form an invariant mass. In figure \ref{fig:mjjsub} (left), we plot the invariant mass of the subleading jets for the 2+1 topology. The QCD distribution is peaked at low values, whereas there is a peak in the signal at high $m_{jj}$. The peak is more pronounced (lower combinatorial background) for low $m_Q$, but also the leakage of QCD events in the distribution is larger.

\paragraph{Gaps:}
\begin{figure}[ht!]
	\begin{center}
		\includegraphics[scale=0.36]{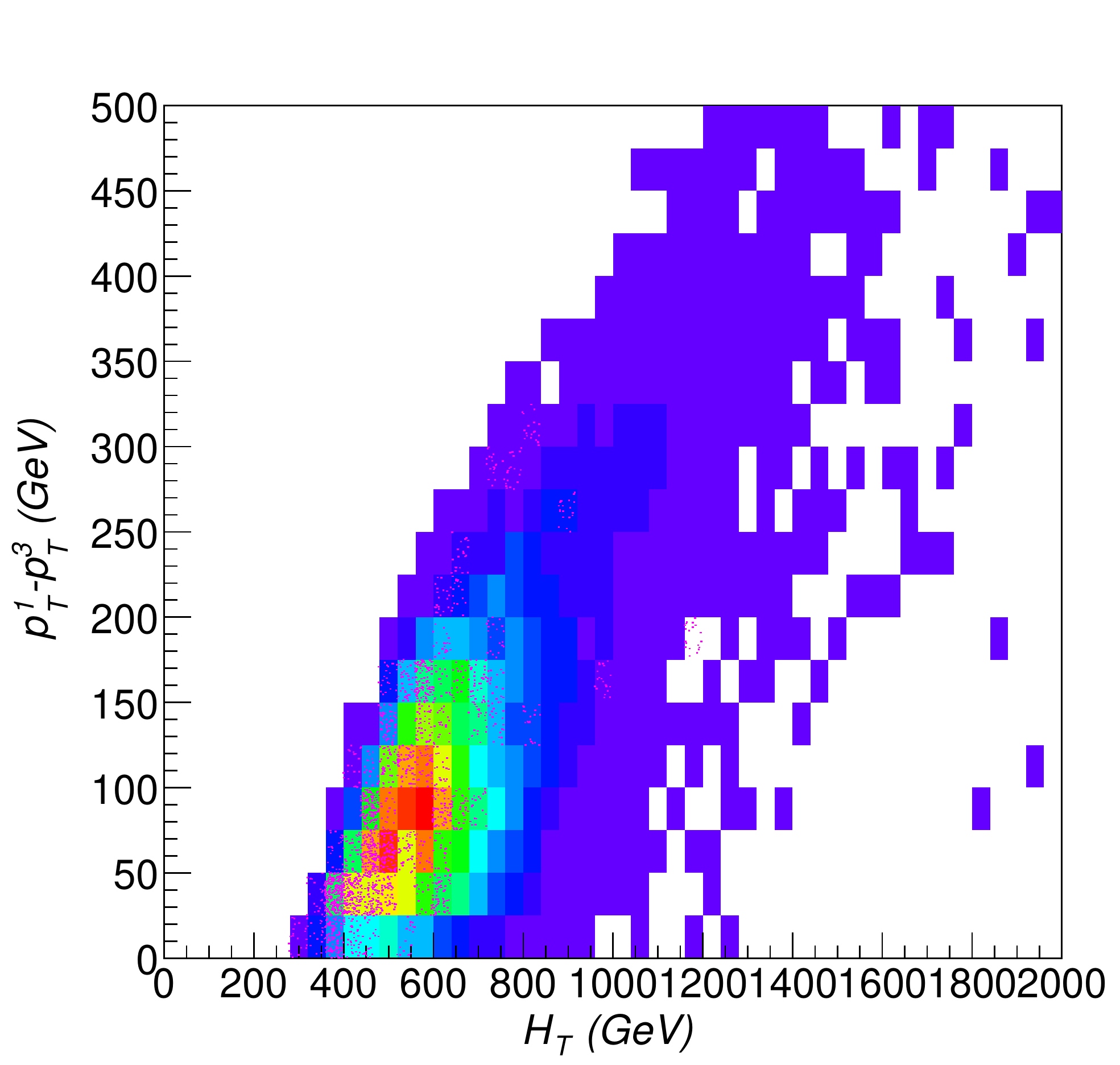}
		\includegraphics[scale=0.36]{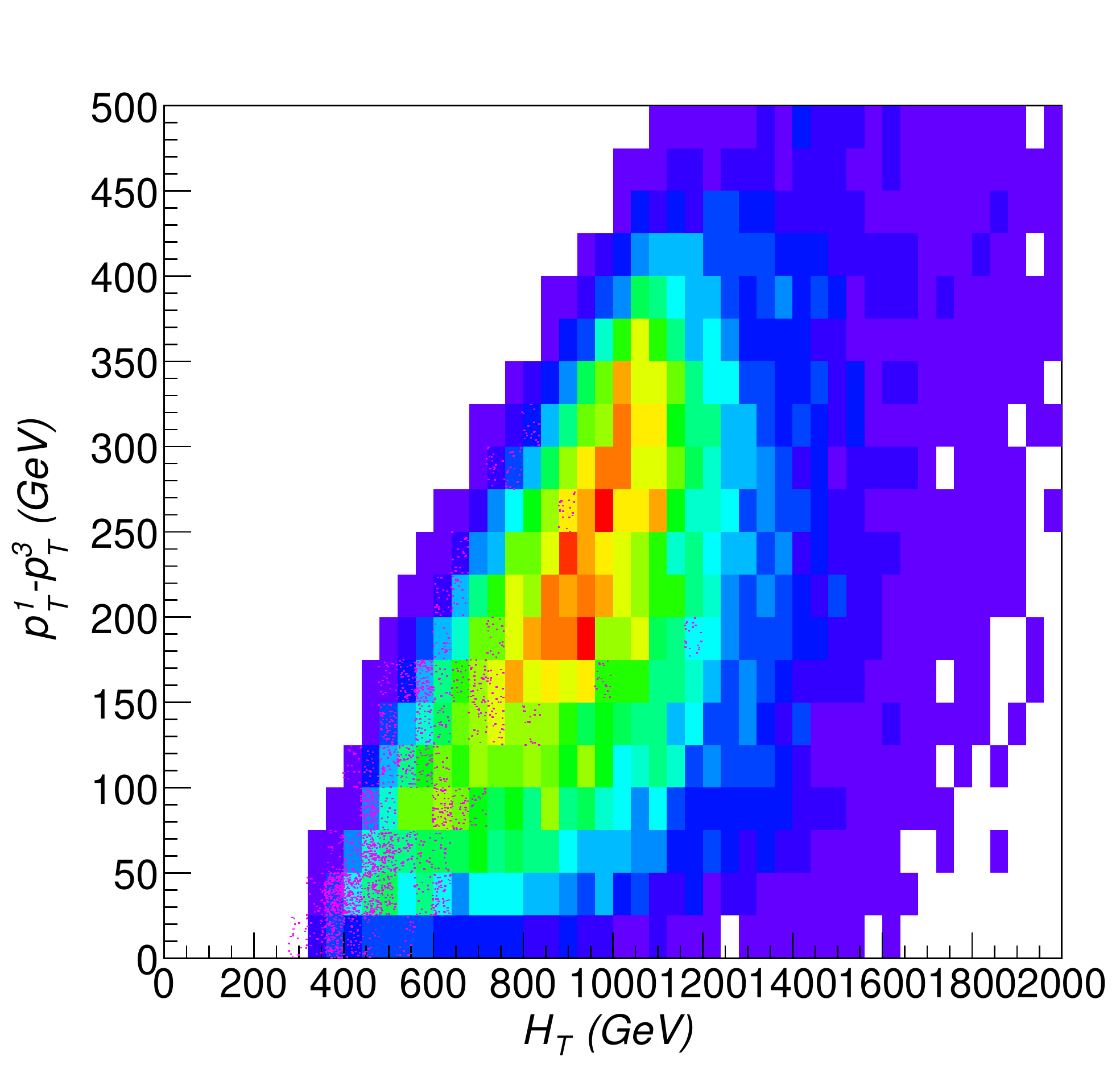}
		\caption{\emph{ $H_T$ versus $p_T^1-p_T^3$ for the 3+1 topology at LHC8 for $m_Q=$ 500 GeV (left) and 1 TeV (right). The pink-scatter plot corresponds to the QCD background. The events are at detector level for LHC8.}}
		\label{fig:abcd}
	\end{center}
\end{figure} \noindent
One could also exploit the gaps among the jets momenta and invariant masses. This is specially interesting in the case of 3+1 topologies, on which we focus in the following. In figure \ref{fig:mjjsub} (right) we plot the variable $\Delta_{24}=\sum_{i=2,3,4} p_{T,i}-m_{234}$, a gap between the $p_T$'s and invariant mass. In the QCD background, the invariant mass and the $p_T$ sum are close to each other, and we expect to be peaked at low values. In the signal events, the invariant mass tends to be smaller than the scalar sum of $p_T$'s.

The gap between the jets in the event can also be used to discriminate between signal and background. In figure \ref{fig:abcd}, we plot $H_T=\sum_{i=1}^4 p_T^i$ versus $\Delta_{13}=p_{T}^1-p_{T}^3$, the $p_T$ difference between the first and third jet, for $m_Q$= 500 GeV (left) and 1 TeV (right). The signal  is characterized by a larger $H_T$ and also by a larger hierarchy between the first and third jet. The differences between QCD and signal are weaker at low $m_Q$, and a harder cut on both variables should be done to keep QCD under control. Although the two variables are clearly correlated, a modified ABCD method could be used here to estimate the amount of QCD background leaking into the signal region.

\paragraph{Obtaining $S/B = 1$:}
We would like to quantify the effect of the cuts on signal and QCD background using the variables described above. In table \ref{tab:cutflow} we describe the cut-flow of those variables for the 2+1 case. The 3+1 case behaves very similarly in terms of signal efficiencies. Note that the QCD background of $n_j \gtrsim 3$, 4 jets with $p_T>$ 70 GeV and $|\eta_j|<$ 2.5 at LHC8 is 3 $\times 10^{4}$ pb and  3 $\times 10^{3}$ pb, respectively. The signal cross section can be read in figure \ref{fig:fermionprodcrossx}  for specific values of $g_{\rho}$, $\sin \phi_{R_{u,d}}$, and it typically varies between 1 to 10 pb for $m_{\rho}\lesssim$  2.5 TeV. To achieve $S/B\sim 1$, one would need to have a relative suppression of efficiencies of $10^{2}-10^{4}$. In the table \ref{tab:cutflow}, one can see how this can be achieved by implementing cuts on the variables described above.
\begin{table}[ht]
	\begin{center}
		\begin{tabular}{|c|c c|c c|}
			\toprule[1pt]
			\multirow{2}{*}{Cut-flow} & \multicolumn{2}{c|}{$m_Q = 600$ GeV} & \multicolumn{2}{c|}{$m_Q = 1200$ GeV} \\
			\cmidrule{2-5}
			& signal & QCD & signal & QCD   \\
			\midrule[1pt]
			  $p_T$ leading jet $>450$ GeV & 0.51 & 0.0067 & 0.90 & 0.0067 \\
			  $H_T > m_Q$ & 0.51 & 0.0067 & 0.80 & 0.0015 \\
			  $|m_{jj} - m_Q| <$ (30, 50) GeV & 0.15 & 0.00037 & 0.11 & 2.5$\times 10^{-5}$ \\
			  $\Delta \phi_{jj} > 1.5$  & 0.045 & 9.9 $\times 10^{-5}$ & 0.060 & $2.1 \times 10^{-7}$ \\
			\bottomrule[1pt]
		\end{tabular}
	\end{center}
	\caption{\emph{Cut-flow demonstrating the effect on signal and background of cutting on the variables presented in the text. The numbers correspond to the efficiency to specified set of cumulative cuts. Here $jj$ is the combination of the two subleading jets. For the background, the final numbers represent the cut-flow with either $m_Q = 600$ GeV or $m_Q = 1200$ GeV.}}
	\label{tab:cutflow}
\end{table}

To produce this cut-flow, we took two benchmark masses, $m_Q$= 600 and 1200 GeV, and the 2+1 signature. We chose the 2+1 topology, as it suffers from the largest background, still interesting $S/B$ can be achieved using these cuts. Note that we have not truly optimized the cuts to a specific signal, and the intention of the table is to show that a background reduction in the required range is possible. Note also that we have not made use of the {\it gap} variables in this cut-flow, which could improve the sensitivity of the search.
\begin{figure}[ht]
	\begin{center}
		\includegraphics[scale=0.46]{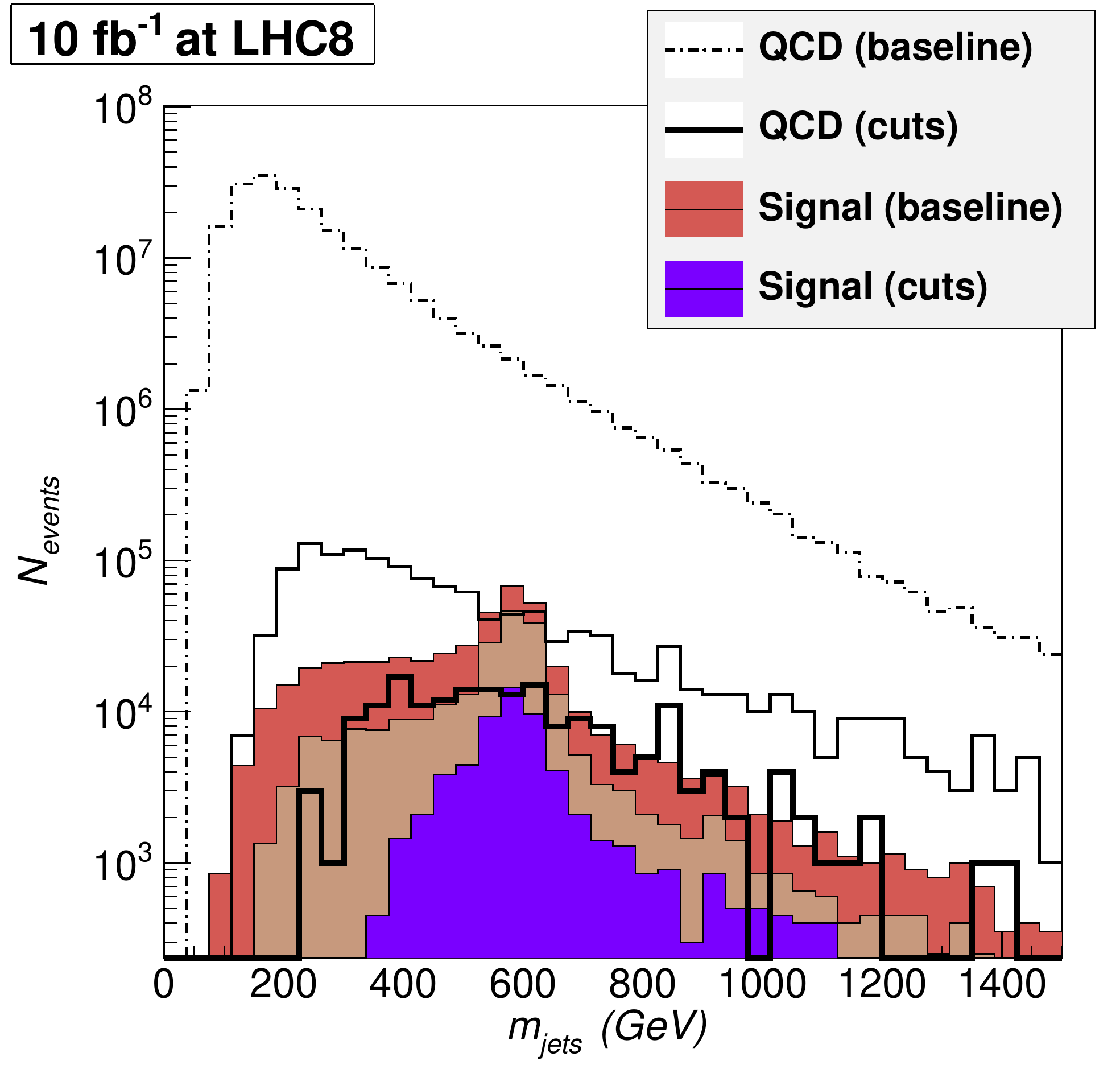}
				\caption{\emph{QCD background (black lines) and signal (solid colors) when the cuts on the table \ref{tab:cutflow} are applied, except the one on $m_{jj}$. See text for details.}}
		\label{fig:cutflow}
	\end{center}
\end{figure}

Note that the cutflow table and figure are produced using detector level events showered with Pythia with MLM matching \cite{bib:MLM} and simulated with Delphes \cite{bib:delphes} with anti-$k_T$ jets of $R=0.7$.  

In figure \ref{fig:cutflow} we illustrate this cut-flow with a normalized background for 10 fb$^{-1}$ of luminosity and a signal of $m_Q=$ 600 GeV and $\sigma=$ 5 pb. In this figure, the three black lines correspond to QCD 3 jets with {\it 1.)} $p_T>$ 70 GeV and $|\eta_j|<$ 2.5, {\it 2.)} $\oplus p_T^{leading}>$ 450 GeV and $H_T> m_Q$ and {\it 3.)} $\oplus  \Delta \phi_{jj}>$ 1.5.
Similarly, the solid histograms correspond to the same cuts, applied now in the signal.

At 14 TeV, the production cross section for QCD with $n_{j}\geqslant$ 3 and $p_T>$ 70 GeV, $|\eta_j|<$ 2.5, increases by a factor three respect to the 8 TeV run. For the $m_Q=$ 600 GeV cut-flow described in table \ref{tab:cutflow}, the efficiencies to pass the cuts increase by a factor $\cal O$(2) from 14 TeV respect to the 8 TeV case. The $m_Q$=1200 GeV is more dramatic, with an efficiency increase for the QCD case of $\gtrsim\cal O$(10). As we already mentioned, the cut-flow presented here should be seen as indicative of the strategy to follow, and it is clear one would need to re-optimize when moving from the 8 TeV to the 14 TeV run.

\section{Conclusions}
\label{sec:conclusions}
In this paper we have investigated the experimental signatures and bounds of partially composite Higgs models where right-handed quarks are strongly
composite. This scenario, strongly motivated by flavor physics, was until recently very weakly constrained experimentally. The situation is rapidly changing with the LHC results that are progressively carving out significant regions of parameter space. We presented  the most relevant bounds that can be extracted from the latest LHC data. It is worth to emphasize that the experimental strategies to test these models at LHC are of a different nature from the ones of the more studied anarchic scenarios or supersymmetry. In particular they typically produce jet final states without leptons or missing energy. For this reason existing analysis are in some cases not optimal and could be improved with dedicated searches.

One of the most important constraints on right-handed quark compositeness arises from dijet searches. These place a direct bound on the spin one gluon resonances. In some regions of parameters these states are excluded up to 3 TeV but the result is strongly sensitive on the fermionic spectrum.

We also derive bounds on the masses of the lightest fermionic partners. These are particularly relevant given their role for the naturalness of the theory. In the light of the 125 GeV Higgs discovery some fermions should be lighter than 1 TeV for a small tuning of the theory \cite{tuning}. One interesting experimental feature is that single production of the new fermions dominates the bounds unlike the case of anarchic scenarios where at present double production produces the strongest constraints.

We derive an extremely strong bound on the left partners that are excluded up to 2 TeV in theories that realize MFV. This is obtained from single electroweak production of partners of the up quark studied by the ATLAS collaboration. Right-handed quark partners can  be singly produced through the gluon resonances with smaller cross sections and different final states. The direct bound is much weaker in this case. Overall our study shows that models that realize MFV are at least as tuned as the anarchic scenarios. This can be avoided abandoning MFV in favour of theories based $SU(2)$ flavor symmetry \cite{bib:su2} where the light generations can be more elementary than the top, see also \cite{bib:Barbieri2012tu} for a related discussion.

We conclude by noting that our scenario motivates more general experimental searches than the ones presently published. This is already possible with the existing data with minor modifications of experimental analyses. In particular our multi-jet signals could be more efficiently captured with a different ordering of jets. Dijet studies should also be extended to trijets. Let us also mention that the multi-jet signals originate from quarks, whereas the background is dominated by high-multiplicity QCD gluons. Therefore, jet tagging techniques, such as in reference \cite{jet-tagging}, would be valuable to reject the background. We hope that these efforts will be pursued by the experiments.

\vspace{0.5cm}
\noindent {\bf Acknowledgments:}
We thank Georges Azuelos, Andrea Banfi, Gilad Perez, Riccardo Rattazzi, and Gavin Salam for useful discussions. MR would like to thank the Galileo Galilei Institute (GGI) in Florence for hospitality while part of this work was carried out. The work of MR is supported in part by the MIUR-FIRB grant RBFR12H1MW. The work of VS is supported by the Science Technology and Facilities Council (STFC) under grant number ST/J000477/1. The work of MdV and AW was supported in part by the German Science Foundation (DFG) under the Collaborative Research Center (SFB) 676.

\appendix

\section{Right-Handed Composite Model}
\label{sec:compositemodel}

In this appendix we describe the effective Lagrangian used in our simulations. This is a simple extension of \cite{bib:2sitescontino}. We will focus on the quark sector. The composite states are multiplets of the global symmetry $SU(3)\times SU(2)_L\times SU(2)_R\times U(1)_X$. As described in section \ref{sec:compositelightquarks} we take the quark partners in the following representations
\begin{eqnarray}
	& \tilde{U} ={\bf(1,1)_{\frac 2 3}} \qquad & \tilde{D} ={\bf(1,1)_{-\frac 1 3}} \nonumber \\
	& L_U = {\bf (2,2)_{\frac 2 3}} = \begin{pmatrix} U & U_{\frac 5 3} \\	D & U_{\frac 2 3} \end{pmatrix} \qquad
	& L_D = {\bf (2,2)_{-\frac 1 3}} = \begin{pmatrix} D_{-\frac 1 3} & U \\	D_{-\frac 4 3} & D \end{pmatrix} ,
\end{eqnarray}
all fundamentals of $SU(3)$. Focusing on the first generation we consider the following Lagrangian for the composite fermions
\begin{align} \label{eq:2sitecustodial}
	\mathcal{L}_\mathrm{composite} = & - \frac{1}{4} \rho_{\mu\nu}^{i2} + \frac{m_\rho^{i2}}{2} \rho_\mu^{i2} + \mathrm{Tr} \left[ |D_\mu H|^2 \right] - V(H) \nonumber \\
	+ & \mathrm{Tr} \left[ \bar{L}_U (i \ddslash{ D}-  m_{L_U}) L_U \right] +\bar{\tilde{U}} (i \ddslash{ D} -  m_{\tilde{U}}) \tilde{U} \nonumber \\
	+ & Y_U \mathrm{Tr} \left[ \bar{L}_{U} \mathcal{H} \right]_L U_R+ \mathrm{h.c.}\nonumber \\
	+ & \left\{ U \to D \right\} .
\end{align}
We only include the composite Yukawas that are relevant for the generation of the SM flavor structure. Among the spin-1 resonances we consider a massive octet of $SU(3)$ and assume that interacts as a gauge field. The elementary Lagrangian is just QCD with massless quarks
\begin{equation} \label{eq:elementarylagrangian}
	\mathcal{L}_\mathrm{elementary} = - \frac{1}{4} G_{\mu\nu}^a G^{a\mu\nu} + \bar{q}_L i \dslash{D} q_L + \bar{u}_R i \dslash{D} u_R + \bar{d}_R i \dslash{D} d_R .
\end{equation}
SM quarks mix with the fermions of equal quantum numbers
\begin{equation} \label{eq:mixingcustodial}
	\mathcal{L}_\mathrm{mixing} =\Delta_{Lu} \bar{q}_L Q_{Ru}+\Delta_{Ru}  \bar{q}_L Q_{Rd}\,+ \Delta_{Ru} \bar{\tilde{U}}_{L} u_{R}  + \Delta_{Rd} \, \bar{\tilde{D}}_{L} d_{R} + \mathrm{h.c.}
\end{equation}
where the $Q_{Ru}$ and $Q_{Rd}$ are the doublets contained in $L_U$ and $L_D$ respectively. We will assume  $\lambda_{Ld} \ll \lambda_{Lu}$. Similarly the gauging of SM symmetries introduces a linear mixing between the SM and the composite spin-1 resonances.

Diagonalizing the elementary-composite mixings  the Lagrangian in the mass basis reads
\begin{align} \label{eq:gaugelagrangian}
	\mathcal{L}_\textrm{gauge} = & - \frac{1}{4} G_{\mu\nu} G^{\mu\nu} + \frac{1}{2} \left( D_\mu \rho_\nu D_\nu \rho_\mu - D_\mu \rho_\nu D_\mu \rho_\nu \right) + \frac{M^2}{2 \cos^2 \theta} \rho_\mu \rho^\mu \nonumber \\
	& + \frac{i g_s}{2} G_{\mu \nu} \left[ \rho_\mu , \rho_\nu \right] + 2 i g_s \cot 2 \theta D_\mu \rho_\nu \left[ \rho_\mu , \rho_\nu \right] + \frac{g_s^2}{4} \left( \frac{\sin^4 \theta}{\cos^2 \theta} + \frac{\cos^4 \theta}{\sin^2 \theta} \right) \left[ \rho_\mu , \rho_\nu \right]^2 ,
\end{align}
for the bosons and
\begin{align} \label{eq:fermionlagrangian}
	\mathcal{L}_\textrm{fermion} = & \, \bar{q}_L i \dslash{D} q_L + \bar{Q}_u \left( i \dslash{D} - m_{Q_u} \right) Q_u  \nonumber \\
	& \, + g_s \bar{q}_L \left( \sin^2 \phi_{Lu} \cot \theta - \cos^2 \phi_{Lu} \tan \theta \right) \rho_\mu \gamma^\mu q_L \nonumber \\
	& \, + g_s \bar{q}_L \left( \frac{\sin \phi_{Lu} \cos \phi_{Lu}}{\sin \theta \cos \theta} \right) \rho_\mu \gamma^\mu Q_{Lu} + \; \mathrm{h.c.} \nonumber \\
	& \, + g_s \bar{Q}_{Lu} \left( \cos^2 \phi_{Lu} \cot \theta - \sin^2 \phi_{Lu} \tan \theta \right) \rho_\mu \gamma^\mu Q_{Lu} \nonumber \\
	& \, + \left\{ (q_L, Q_u) \to (q_L, Q_d) \; , (u_R, U_L) \; , (d_R, D_L) \right\} .
\end{align}
for the fermions. In the expressions above $\tan \theta = g_\mathrm{el}$,  $g_\rho$ $\tan \phi = \Delta / m$  and $g_s = g_\mathrm{el} \cos \theta$ is the QCD coupling. This is the final form of the Lagrangian which has been implemented in FeynRules \cite{bib:feynrules} to study the LHC phenomenology.

The Higgs vacuum expectation value introduces the following mixings in the up sector
\begin{equation}
	\mathcal{L}_\mathrm{int}^L = - \frac {Y_U v}{\sqrt{2}}  \sin \phi_{Ru}  \bar{u}_R \left[ U+ U_{2/3} \right] + \mathrm{h.c.}
\end{equation}
Diagonalizing these terms generates the electroweak interactions of equation \eqref{eq:electroweak} relevant for single production
of left-handed partners \cite{bib:singleproduction,bib:azuelos}.

\begin{figure}[ht]
	\begin{center}
		\includegraphics[scale=0.3]{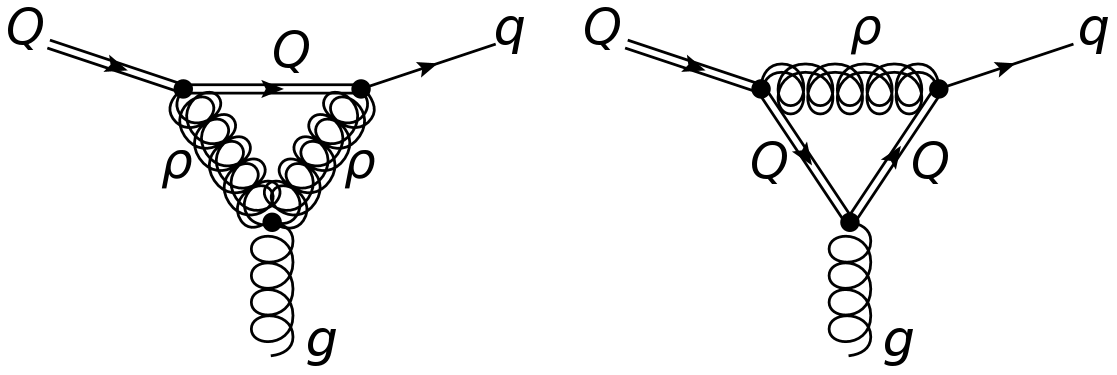} \hspace{.2cm}
		\caption{One-loop new physics contributions to the chromomagnetic operator in partially composite models.}
		\label{fig:chromodiagrams}
	\end{center}
\end{figure}
Finally in the effective Lagrangian of the strong sector we include the dimension 5 operator
\begin{equation}
	\mathcal{L}_\mathrm{chromo} = \frac {g_s\,\kappa_0 }{m_Q}\,  \bar{Q}_L \sigma_{\mu\nu} T^a q_R G_{\mu\nu}^a + \mathrm{h.c.}
\end{equation}
The chromomagnetic interaction is generated by loops of the strong sector fields, see figure \ref{fig:chromodiagrams}. The naive estimate is $\kappa_{0}  \sim \frac {g_\rho^2}{16 \pi^2}$. We will however be interested in the region $m_Q< m_\rho$. In this case the loops generate
\begin{equation}
\kappa_0 \sim \frac{g_\rho^2}{16 \pi^2} \frac{m_Q^2}{m_\rho^2} .
\end{equation}
Dressing the operator with the mixing the interaction \eqref{eq:chromo} is obtained. In our numerical evaluation we will use the estimate
\begin{equation}
	\kappa = \frac{N_c}{32 \pi^2} \frac{m_Q^2}{m_\rho^2} X_{R}^{QQ} X_{R}^{Qq} .
\end{equation}
The suppression is relevant phenomenologically because it renders two body and three body decay widths comparable.

\section{Approximate \texorpdfstring{$p_T$}{pT} Distribution}
\label{sec:ptordering}
In the following, we derive the approximate cross section given in equation \eqref{eq:dsigmadpT} for the process $u \, u \to u \, U$. We only include the relevant left-handed couplings, a good approximation for $g_\rho\gg g$ and derive the $t$-channel expression\footnote{There is a an additional $u$-channel contribution but since $\Delta y \to 0$ minimizes $\hat s$, see \eqref{eq:stuQ}, we typically have $t \approx u$.} for a given $p_T$ of the spectator quark. The amplitude squared summed over initial and final states is proportional to
\begin{equation}
\left| \overline{\mathcal{M}_{fi}} \right |^2 \propto \frac{\hat s ( \hat s  - m_Q^2)}{(\hat t - m_\rho^2)^2} .
\end{equation}
This leads to a dependence on the Mandelstam variables in the cross section given by 
\begin{equation}
\frac{d^3\sigma}{dy_3 dy_4 d|p_T|} \propto  (f(x_1) \, f(x_2))\,\frac{p_T}{S} \, \frac{ \hat s  - m_Q^2}{(\hat t - m_\rho^2)^2} ,
\end{equation}
where $y_{3,4}$ are the rapidities of the daughter particles, $S$ is the c.o.m. energy and $x_{1,2}$ are the usual partonic momentum fractions carried by the initial partons. Recalling that
\begin{align}\label{eq:stuQ}
\hat s &= m_Q^2 + 2\, p_T^2 + 2\, p_T \sqrt{m_Q^2 + p_T^2} \cosh\Delta y \\
\hat t &= -p_T \left( p_T + \sqrt{m_Q^2 + p_T^2}\, \exp{(-\Delta y)}\right) \\
\hat u &= -p_T \left( p_T + \sqrt{m_Q^2 + p_T^2}\, \exp{(\Delta y)}\right),
\end{align}
we find
\begin{equation}
\frac{d^3\sigma}{dy_3 dy_4 d|p_T|} \propto (f(x_1) \, f(x_2))\,\frac{p_T^2 \left(\cosh (\Delta y)
   \sqrt{m_Q^2+p_T^2}+p_T\right)}{36 \pi
   \left( m_\rho^2+p_T \left(e^{-\Delta y}
   \sqrt{m_Q^2+p_T^2}+p_T\right)\right)^2} .
\end{equation}
\begin{figure}[ht]
	\begin{center}
		\includegraphics[scale=0.85]{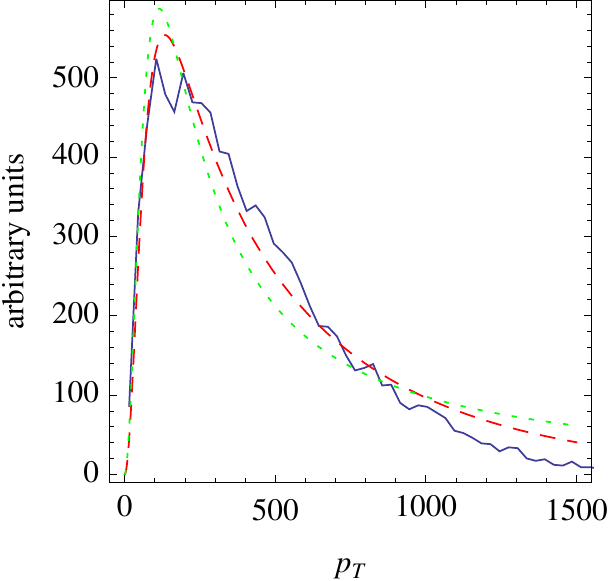} \quad 
		\includegraphics[scale=0.85]{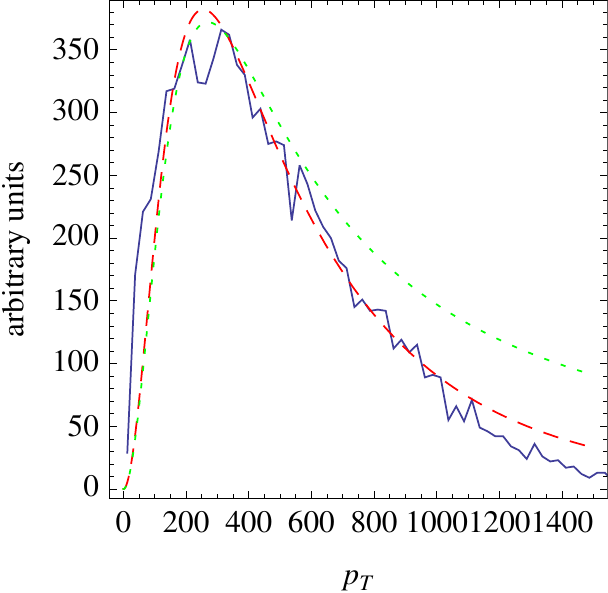} \quad 
		\includegraphics[scale=0.85]{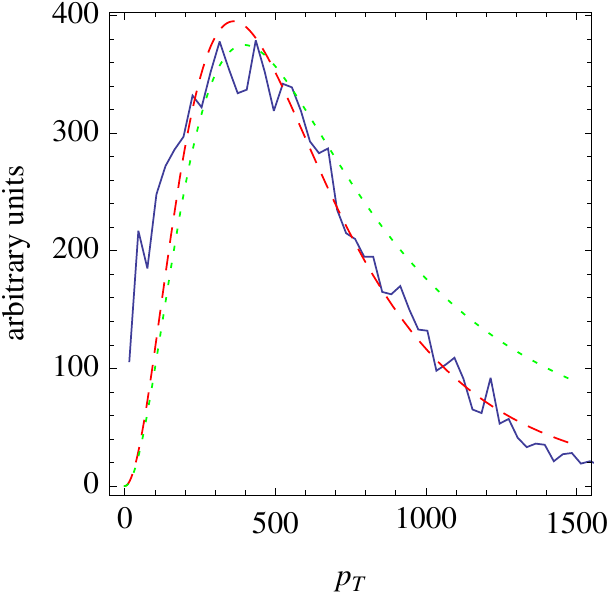}
		\caption{\emph{\label{fig:ptapprox} Comparison between approximate and exact $p_T$ distributions for various values of the fermionic partner and octet masses $(m_Q , m_\rho) =  (100 , 2500) ; \; (250 , 2500); \; (500 , 2500)$ with $g_\rho = 6$ and $\sin \phi_{R_{u,d}} = 0.6$. The red line uses the full $t$-channel propagator, whereas the green line assumes a contact interaction. We use constant $\alpha$ in the plots. Naturally, for very large $p_T$ the $\rho$ dynamics is resolved and taking $\alpha$ constant ceases to be a good approximation.}}
	\end{center}
\end{figure}
We can now derive a simple approximation for the $p_T$ distribution. Since $\Delta y \to 0$ minimizes $\hat s$, we can set  $\Delta y =0$ in the following. The parton luminosities are steeply falling functions of $\hat s/S$, therefore we can approximate the remaining integration by the threshold value of the parton luminosities which we model as a steeply falling polynomial $(\hat s/S)^{-\alpha}$. We extracted $\alpha$ from the MSTW2008 pdfs \cite{Martin:2009iq}. For heavy color octets ($m_\rho \gg m_Q, p_T$), we can also ignore the octet propagator. Combining these approximations we find
\begin{equation}
\frac{d\sigma}{d|p_T|} \propto \frac{1}{S}\, \frac{p_T^2}{
   m_\rho^4} \left(p_T+ \sqrt{m_Q^2+p_T^2}\right) \left( \frac{ p_T^2+m_Q^2+ p_T\sqrt{m_Q^2+p_T^2}}{S}\right)^{-\alpha}
\end{equation}
where $\alpha \sim 3-6 $ is a slowly varying function of $\hat s$ determined by the parton luminosities. The maximum of the $p_T$ distribution is therefore approximately at $(p_T)^{\rm max} \approx \frac{1.5}{\sqrt{4\alpha -6}} m_Q \approx \frac12\, m_Q $. In figure \ref{fig:ptapprox} we compare the above approximation with a parton level simulation using the full model implementation, validating the result and the approximations.

\bibliographystyle{unsrt}

\end{document}